\documentclass[11pt,a4paper]{article}

\usepackage{slashed}
\usepackage[hidelinks]{hyperref}
\usepackage{amsthm}
\usepackage{mathtools}
\usepackage{graphicx}
\usepackage{amssymb}
\usepackage{amsmath}
\usepackage{amsfonts}
\usepackage{simplewick}
\usepackage{tikz-cd}
\usetikzlibrary{decorations.markings}
\usetikzlibrary{intersections}
\usetikzlibrary{calc}
\usepackage{latexsym,cite}
\usepackage{bbm}
\usepackage{float}

\usepackage{mathrsfs, mathtools, xpatch}
\usepackage{dsfont}

\usepackage{euscript}

\usepackage[shortlabels]{enumitem}

\newcommand{\fra}{\mathfrak{a}}
\newcommand{\frso}{\mathfrak{so}}

\def\swone{{\textrm{\tiny $(1)$}}}

\def\swzero{{\textrm{\tiny $(0)$}}}

\def\1PI{{\textrm{\tiny 1PI}}}

\newcommand{\id}{\mathrm{id}}
\newcommand{\CF}{\mathcal{F}}

\newcommand{\CCV}{\mathscr{V}}
\newcommand{\CCR}{\mathscr{R}}
\newcommand{\CCC}{\mathscr{C}}
\newcommand{\CR}{\mathcal{R}}
\newcommand{\CA}{\mathcal{A}}
\newcommand{\ii}{{\mathrm{i}}}

\newcommand{\sfR}{\mathsf{R}}

\newcommand{\nn}{\nonumber}
\def\d{{\rm d}}
\def\ds{\stackrel{\star}{,}}

\def\RR{{\mathcal R}}

\newcommand{\mbf}[1]{{\boldsymbol {#1} }}

\newcommand{\sfh}{{\sf h}}
\newcommand{\sfp}{{\sf p}}
\newcommand{\sfm}{{\sf m}}
\newcommand{\sfDelta}{{\mathsf{\Delta}}}
\newcommand{\sfLambda}{{\mathsf{\Lambda}}}
\newcommand{\sgreen}{\mathsf{G}}
\newcommand{\sP}{\mathsf{P}}
\newcommand{\sH}{\mathsf{H}}
\newcommand{\sI}{\mathsf{I}}
\def\swone{{\textrm{\tiny $(1)$}}}

\newcommand{\Sym}{\mathrm{Sym}}
\newcommand{\RZ}{\mathbbm{Z}}
\newcommand{\FR}{\mathbbm{R}}
\newcommand{\FC}{\mathbbm{C}}
\newcommand{\BVL}{{\Delta}_{\textrm{\tiny BV}}}
\newcommand{\MC}{{\textrm{\tiny MC}}}
\newcommand{\CS}{\mathcal{S}}
\newcommand{\tte}{\mathtt{e}}
\newcommand{\ttc}{\mathtt{c}}
\newcommand{\ttv}{\mathtt{v}}
\newcommand{\ttf}{\mathtt{f}}

\newcommand{\ttF}{\mathtt{F}}
\newcommand{\e}{{\mathrm{e}}}
\newcommand{\dd}{\mathrm{d}}
\newcommand{\im}{\mathrm{i}}

\definecolor{dgreen}{rgb}{.2,.6,.2}

\definecolor{dred}{RGB}{250,0,100}

\numberwithin{equation}{section}
\catcode`@=12

\newenvironment{myitemize}{\begin{itemize}[itemsep=0.1cm, leftmargin=*, topsep=0cm]}{\end{itemize}}

\begin{document}
 
\renewcommand{\thefootnote}{\fnsymbol{footnote}}

\vspace{1cm}

\begin{center}

{\LARGE{\bf Batalin-Vilkovisky quantization with an angular twist}}

\vspace*{1.5cm}

\baselineskip=14pt
		
{\large\bf Djordje Bogdanovi\'c}${}^{\,(a)\,,\,}$\footnote{Email: \ {\tt
djbogdan@ipb.ac.rs}} \ \ \ \ \ {\large\bf Marija Dimitrijevi\'c \'Ciri\'c}${}^{\,(a)\,,\,}$\footnote{Email: \ {\tt
dmarija@ipb.ac.rs}} \ \ \ \ \ {\large\bf Richard J. Szabo}${}^{\,(b)\,,\,}$\footnote{Email: \ {\tt r.j.szabo@hw.ac.uk}} 
\\[6mm]

\noindent  ${}^{(a)}$ {\it Faculty of Physics, University of
Belgrade}\\ {\it Studentski trg 12, 11000 Beograd, Serbia}
\\[3mm]

\noindent  ${}^{(b)}$ {\it Department of Mathematics, Heriot-Watt University\\ Colin Maclaurin Building,
Riccarton, Edinburgh EH14 4AS, U.K.}\\ and {\it Maxwell Institute for
Mathematical Sciences, Edinburgh, U.K.}
\\[30mm]

\end{center}

\begin{abstract}
\noindent
We construct cubic scalar field theory on $\lambda$-Minkowski space by combining the Batalin-Vilkovisky formalism with harmonic analysis, and produce two inequivalent noncommutative quantum field theories. The braided theory is based on a braided $L_\infty$-algebra whereby covariance dictates a spectral decomposition into cylindrical Bessel functions that diagonalise the angular Drinfel'd twist; in this theory we find the usual logarithmic ultraviolet divergences and confirm the absence of UV/IR mixing. The standard noncommutative theory is based on a classical $L_\infty$-algebra; in this theory we relate the spectral decompositions into plane wave and cylindrical harmonic eigenmodes of the Klein-Gordan operator, we verify the planar equivalence theorem, and we demonstrate a periodic form of UV/IR mixing in which non-planar correlators are generically ultraviolet finite but become non-analytic on an infinite lattice of exceptional momenta.
\end{abstract}

\vspace*{2cm}

\noindent
{\small{\bf Keywords:} {angular twist, $\lambda$-Minkowski space, scalar field theory, BV quantization}} \normalsize


\newpage

{\baselineskip=10pt
\tableofcontents
}

\baselineskip=14pt

\setcounter{footnote}{0}
\renewcommand{\thefootnote}{\arabic{footnote}}

\bigskip

\newcommand{\Section}[1]{\setcounter{equation}{0}\section{#1}}
\renewcommand{\theequation}{\arabic{section}.\arabic{equation}}

\section{Introduction}%

Noncommutative field theories have been of interest for decades due to their role as effective theories spanning condensed matter physics to string theory as well as other models of quantum gravity (see e.g.~\cite{Douglas:2001ba, Szabo:2001kg, Hersent:2022gry,Wallet:2025xbp, Szabo:2025mxr} for reviews). As quantum field theories they have introduced new concepts and revealed novel physical phenomena. One of their most intriguing and at the same time pathological features is ultraviolet/infrared (UV/IR) mixing: high-energy divergences reappear as low-energy divergences in higher-loop diagrams. UV/IR mixing was first observed in \cite{Minwalla:1999px} in the context of scalar field theories with Moyal noncommutative deformation, and was subsequently discussed in more intricate models of noncommutative spacetimes. Open problems include conceptual issues surrounding the renormalizablity of scalar field theories and the construction of renormalizable  gauge theories in the star-product formalism.

The Drinfel'd twist deformation formalism  enables the construction of broad classes of noncommutative field theories through natural definitions of star-products, noncommutative differential calculi and deformed symmetries. Those twists which are based on the universal enveloping Hopf algebra of the Poincar\'e algebra are of particular interest due to their role in integrable deformations of the AdS/CFT correspondence~\cite{vanTongeren:2015uha,vanTongeren:2016eeb} and the fact that, among other notable features, they provide a well-defined Hodge dual operation necessary for constructing Yang-Mills theories \cite{Meier:2023lku}. In this paper we will revisit the quantization and UV/IR mixing problem for an example of a Drinfel'd twist deformed scalar field theory from a new perspective. Our approach is rooted in a novel combination of two techniques: the Batalin-Vilkovisky (BV) formalism and harmonic analysis.

The most studied examples of noncommutative field theories are defined with the Moyal star-product. They can be obtained via a twist constructed from the generators $P_\mu$ of translations on $\FR^d$ as
\begin{equation}
\mathcal{F}_\text{Moyal}=\exp\big(\tfrac{\ii}{2}\,\theta^{\mu\nu}\,P_{\mu}\otimes P_{\nu}\big) \ ,  \nn
\end{equation}
where $(\theta^{\mu\nu})$ is a real $d{\times}d$ constant antisymmetric matrix. The Moyal twist defines a noncommutative deformation of Minkowski space $\FR^{1,d-1}$ and can be used to deform the usual pointwise product of functions on $\FR^d$ to a star-product $\star$, which leads to the canonical (constant) noncommutativity between coordinate functions
\begin{equation}
[x^\mu \ds x^\nu] := x^\mu\star x^\nu - x^\nu\star x^\mu = \ii\,\theta^{\mu\nu} \ .\nn
\end{equation}

The angular twist \cite{Ciric:2017rnf} is a straightforward generalisation of the Moyal twist on $\FR^4$ using a pair of commuting generators of rotations in a spatial plane $\FR^2\subset\FR^4$ and translations in a normal direction $\FR\subset\FR^4$ to the plane. For a spatial normal direction the corresponding spacetime is called $\lambda$-Minkowski space and it 
represents a noncommutative deformation of Minkowski space $\FR^{1,3}$ in which translational invariance is broken in the plane. Denoting Cartesian coordinates of $\FR^4$ as $(x^\mu) = (t,x^i)=(t,x,y,z)$, where $t$ is time, the noncommutative relations among coordinate functions are now
\begin{equation}
[t \ds x^i] =0 \quad , \quad [z \ds x] = -\ii\,\lambda\, y \quad , \quad [z\ds y] = \ii\,\lambda\, x \, \nn
\end{equation}
for a constant $\lambda\in\FR$. The closely related $\rho$-Minkowski space interchanges the roles of $t$ and $z$, and renames the deformation parameter $\lambda$ to $\rho$. 

Noncommutative scalar $\varPhi^4$-theory based on the angular twist is studied in \cite{DimitrijevicCiric:2018blz, Hersent:2023lqm, Wallet:2025mbv}. Compared to the Moyal deformed $\varPhi^4$-theory on $\FR^{1,3}$, noncommutativity improves the ultraviolet behaviour of non-planar diagrams, reducing quadratic divergences to logarithmic divergences. However, UV/IR mixing persists and the theory remains non-renormalizable. An alternative quantization scheme, which is covariant under the $\rho$-Poincar\'e group, was proposed in~\cite{Fabiano:2025pub} and shown to eliminate UV/IR mixing on $\rho$-Minkowski space. This is reminescent of recent developments which interpret noncommutative field theories as braided field theories~\cite{DimitrijevicCiric:2021jea,Giotopoulos:2021ieg}, whose quantization in the case of Moyal deformed $\varPhi^4$-theory~\cite{DimitrijevicCiric:2023hua} as well as $\varPhi^3$-theory~\cite{Bogdanovic:2024jnf} likewise bypasses the UV/IR mixing problem. One of the main goals of the present paper is to develop braided scalar  quantum field theories with the angular twist deformation, as the first example that goes beyond the conventional Moyal twist deformation, and to provide a detailed side-by-side comparison with the more standard noncommutative scalar quantum field theories for the first time. 

The construction of noncommutative quantum field theories advocated in \cite{Nguyen:2021rsa, Bogdanovic:2024jnf} is based on the modern algebraic approach to the BV formalism due to \cite{Costello:2021jvx}. The great virtue of this approach is that it does not introduce a path integral and instead gives an algebraic prescription for calculating correlation functions. This is particularly relevant for noncommutative field theories where it is difficult to consistently define a path integral: among other issues, it is not obvious what the measure on the space of noncommutative fields should be. Although string theory predicts that the standard Feynman measure should be used, this generically leads to the fatal non-renormalizability issues discussed above and moreover is not compatible with the deformed symmetries of noncommutative spacetime. 

Braided quantum field theory is based on a braided generalisation of BV quantization, typically (but not exclusively) through a Drinfel'd twist deformation of a classical BV theory. By construction these theories are covariant under a noncocommutative triangular Hopf algebra of symmetries; in particular, the space of noncommutative fields becomes an object in the symmetric monoidal category of representations (modules) of the Hopf algebra. This in turn leads to a deformation of  Wick's theorem to a braided Wick theorem and a deformation of the interaction vertex to a noncommutative interaction vertex.
Forgetting the module structures, and hence resorting to the trivial braiding provided by the usual flip map on the category of vector spaces, one recovers the standard Feynman path integral quantization of the noncommutative field theory with the usual undeformed Wick theorem and the noncommutative interaction vertex. In this latter approach the noncocommutative triangular Hopf algebra of symmetries is not fully implemented.

In the case of the Moyal twist this braided BV quantization eliminates non-planar diagrams and hence UV/IR mixing, so that the noncommutative contributions to higher-loop diagrams amount simply to phase factors in external momenta. The resulting noncommutative field theories are thereby renormalizable in the standard textbook sense, but they still fail to provide an ultraviolet completion of quantum field theory. This was first discussed in~\cite{Oeckl:2000eg} from a different algebraic approach to braided quantum field theory, and subsequently rediscovered from various points of view, see e.g.~\cite{Balachandran:2005pn,Grosse:2007vr}. In this paper we explore whether this feature continues to hold beyond the Moyal deformation by constructing braided scalar $\varPhi^3$-theory on $\lambda$-Minkowski space.

Compared to the traditional approach, braided BV quantization comes with a proliferation of braiding operators. For the Moyal twist, these operators are diagonalised in the standard basis of plane waves for the Fourier decomposition of scalar fields on $\FR^d$, which also diagonalise the free propagator of the theory. However, this is no longer true for the angular twist, and while the standard noncommutative quantum field theory can be nevertheless studied in this basis as in~\cite{DimitrijevicCiric:2018blz,Hersent:2023lqm,Wallet:2025mbv}, in the braided theory the plane wave basis introduces rather severe technical complications and is inconvenient to use. It also violates the covariance requirements of braided BV quantization. 

We circumvent this problem in the following way. On general grounds, any abelian Drinfel'd twist can be locally cast in a Moyal-like form via a suitable change of coordinates~\cite{Aschieri:2009qh}; for the angular twist these are cylindrical coordinates on a spatial section $\FR^3\subset\FR^4$. The special feature of the angular twist is that there exists a basis of cylindrical harmonics in these coordinates that diagonalises both the free propagator and the twist, and consequently all braiding operators. Then all calculations are formally identical to those performed for the Moyal twist, with the caveat that correlation functions are now expressed in terms of radial and angular momenta instead of the more familiar four-momenta in Cartesian coordinates, and involve analytical manipulations with cylindrical Bessel functions. One of the main achievements of this paper is the demonstration of how to reproduce standard features of quantum field theory in this unconventional basis of states. In particular, we illustrate how the usual logarithmic ultraviolet divergences of $\varPhi^3$-theory in four dimensions are reproduced in planar diagrams, as well that non-planar diagrams are eliminated and the braided scalar field theory is again free from UV/IR mixing.

To illustrate the versatility of our new approach, we also treat the standard noncommutative $\varPhi^3$-theory on $\lambda$-Minkowski space for the first time. While neither the usual (unbraided) BV formalism nor path integral quantization require the basis of cylindrical harmonics, we believe that it is nonetheless more natural and convenient to work with. In particular, we demonstrate how the angular twist deformation generically eliminates logarithmic ultraviolet divergences in non-planar diagrams and renders them \emph{finite}. UV/IR mixing now appears in a much more severe form compared to the Moyal deformation (a similar behaviour was also mentioned in passing for the noncommutative $\varPhi^4$-theory in~\cite{DimitrijevicCiric:2018blz}): in the limit of vanishing  angular momentum there is an infinite lattice $\frac{2\pi}\lambda\,\RZ\subset\FR$ of exceptional  axial momenta at which the non-planar diagrams exhibit non-analytic behaviour, in that they diverge at these points and the usual logarithmic ultraviolet divergences of the planar diagrams reappear. This makes the quantum field theory much more pathological than usual, because the axial momentum itself is not a periodic variable. We refer to this phenomenon as `periodic UV/IR mixing'. 

We also quantize the usual BV theory in the conventional plane wave basis and present a detailed comparison with the calculations in the novel basis of cylindrical harmonics. We verify the equivalence of planar contributions to the commutative results (up to symmetry factors and phase factors depending only on external momenta) as in the case of the Moyal twist, which went previously unnoticed in the earlier calculations of~\cite{DimitrijevicCiric:2018blz,Hersent:2023lqm,Wallet:2025mbv}, and we recover precisely the same ultraviolet finite non-planar contributions with periodic UV/IR mixing. We demonstrate the equivalence of correlators computed in the two sets of bases by deriving an explicit Fourier series transformation between them, which also elucidates the physical meaning of the deformed momentum conservation laws in the plane wave basis.

\subsubsection*{Outline}

The content in the remainder of this paper is organized as follows: 
\begin{myitemize}
\item In \underline{\sf Section~\ref{First}} we review the basics of angular twist differential geometry and solve the Helmholtz equation for the Klein-Gordon operator in cylindrical coordinates. Separation of variables reduces the radial equation to the Bessel equation.
This yields the alternative basis of cylindrical harmonics for the spectral decomposition of noncommutative scalar fields that we use throughout this paper, in addition to the standard plane wave basis. 
\item In \underline{\sf Section~\ref{QFT}} we review the algebraic BV quantization techniques based on $L_\infty$-algebras and homological perturbation theory. In the rest of the paper this method is applied to the  $\varPhi^3$-model on $\lambda$-Minkowski space. Depending on the choice of corresponding $L_\infty$-algebra, we obtain two  inequivalent noncommutative quantum field theories: Although the two models have identical classical action, their quantization is different.
\item The braided quantum field theory is discussed in \underline{\sf Section~\ref{BraidedQFT}}; the natural building blocks are in terms of cylindrical Bessel functions of the first kind. It corresponds to the theory with underlying braided $L_\infty$-algebra and explicit noncommutativity given by the braiding determined from the angular twist. Wick's Theorem is deformed to the Braided Wick Theorem. It gives rise to a renormalizable noncommutative $\varPhi^3$-theory. In particular, non-planar diagrams and UV/IR mixing are absent from this model. 
\item The standard noncommutative quantum field theory is discussed in \underline{\sf Section~\ref{StarQFT}}. It corresponds to the theory with undeformed (unbraided) underlying $L_\infty$-algebra and implicit noncommutativity contained in the star-product generated by the angular twist. Wick's Theorem remains undeformed. It reproduces the standard expectations, a non-renormalizable noncommutative $\varPhi^3$-theory with UV/IR mixing in  non-planar diagrams, but now in the much more severe form of \emph{periodic} UV/IR mixing. We perform the calculations in both the cylindrical harmonic and plane wave bases, and derive the explicit transformations of correlation functions between the two bases.
\item We conclude with some comments, discussion and outlook for future directions in \underline{\sf Section~\ref{Outlook}}. 
\item Some useful properties of the cylindical Bessel functions of the first kind employed in the calculations of the main text are summarised in \underline{\sf Appendix~\ref{app:Bessel}}.
\end{myitemize}

\section{Angular twist deformation of Minkowski space} \label{First}%

The Drinfel'd twist formalism allows for various noncommutative deformations of Minkowski space and enables the construction of field theories thereon. Noncommutativity breaks the standard Poincar\'e symmetry and replaces it with a twisted Poincar\'e symmetry. Twisted Poincar\'e symmetry for field theories with the canonical Moyal deformation was introduced in~\cite{Wess:2003da,Chaichian:2004yh}, and elaborated in detail in \cite{Aschieri2009}. The $\kappa$-Minkowski space, originally introduced in \cite{Lukierski:1992dt} as a module over the $\kappa$-deformed Poincar\'e algebra, can also be  realised as a twist deformation. ${\sf U}(1)$ gauge theory on $\kappa$-Minkowski space is studied in~\cite{Dimitrijevic:2014dxa}, while  certain cosmological implications  have been explored via a Jordanian twist formalism in \cite{Borowiec:2008uj,Aschieri:2020yft}. 

The $\rho$- and $\lambda$-Minkowski spaces, which appear as members of a family with twisted Poincar\'e symmetries including $\kappa$-Minkowski space~\cite{Lukierski:2005fc}, arise from an angular twist deformation \cite{Ciric:2017rnf}; quantum field theory and quantum mechanics on these spaces are discussed in \cite{DimitrijevicCiric:2018blz, Gubitosi:2021itz, Lizzi:2022hcq, Hersent:2023lqm, Wallet:2025mbv}. In this section we briefly review the deformation leading to the $\lambda$-Minkowski space, and then present a novel framework for studying scalar field theories thereon in terms of cylindrical harmonics.

\subsection{The angular Drinfel'd twist} \label{Twist}%

The angular twist deformation was originally defined on four-dimensional Minkowski space $\FR^{1,3}$~\cite{Ciric:2017rnf}, with signature $(+---)$, and in the following we work in this simplest setting in order to highlight the essential features of our formalism without too much clutter in the formulas. The analysis readily generalizes to arbitrary dimensions for the angular twists discussed in~\cite[Example~4.27]{Giotopoulos:2021ieg}. We write global Cartesian coordinates on $\FR^4$ as $(x^\mu) = (t,x,y,z)$ where $t$ is the time coordinate. The corresponding holonomic vector fields along these directions are written as $\partial_{x^\mu}=\frac\partial{\partial x^\mu}$.

The angular twist operator is constructed as a formal power series with coefficients in $U\fra\otimes U\fra$, where $U\fra$ is the universal enveloping Hopf algebra of a two-dimensional abelian subalgebra $$\fra:=\FR\oplus\mathfrak{so}(2) \ \subset \ \mathfrak{iso}(1,3)$$ of the Poincar\'e algebra, regarded as a subalgebra of the Lie algebra of vector fields on $\FR^4$. It is given by the bidifferential operator
\begin{equation}
\mathcal{F}:=\exp\big({-\tfrac{\im}{2}\,\varTheta ^{AB}\,X_{A}\otimes X_{B}}\big) \ ,
\label{AngTwistCart}
\end{equation}
where $A,B\in\{1,2\}$ and $\varTheta ^{AB}$ is the $2{\times}2$ constant antisymmetric matrix
\begin{equation}
\varTheta^{AB} = { \small \bigg( {
\begin{matrix}
0 & \lambda \\ 
-\lambda & 0
\end{matrix}
} \bigg) } \normalsize \ ,  \notag
\end{equation}
with $\lambda\in\FR$ a formal deformation parameter. 

The vector fields in \eqref{AngTwistCart} are defined as $$X_{1}=\partial _z \qquad , \qquad
X_{2}= x\,\partial_y - y\,\partial_x \ . $$ These vector fields commute, $[X_{1},X_{2}]=0$, and so the twist (\ref{AngTwistCart}) belongs to the broad class of abelian Drinfel'd twists. The terminology `angular' refers to the fact that the vector field $X_{2}$ generates rotations around the $z$-axis, i.e. it can be represented locally as $X_2 = \partial_\varphi$ where $\varphi$ is the polar angle in the $(x,y)$-plane.

The twist (\ref{AngTwistCart}) deforms the cocommutative Hopf algebra $U\fra$ to a noncocommutative Hopf algebra $U_\CF\,\fra$ with the twisted coproduct $$\sfDelta_\CF = \CF\circ\sfDelta\circ\CF^{-1} \ , $$ where $\sfDelta$ is the primitive coproduct of $U\fra$. It
breaks translational invariance in the $(x,y)$-plane, leading to a deformation of the coproducts of the momentum operators  $P_{x^\mu}=-\im\,\partial_{x^\mu}$ according to
\begin{align}
\begin{split}
&\sfDelta_{\cal F}(P_{t}) = P_{t}\otimes 1 + 1\otimes P_{t} \ , \\[4pt]
&\sfDelta_{\cal F}(P_{z}) = P_{z}\otimes 1 + 1\otimes P_{z} \ , \\[4pt]
&\sfDelta_{\cal F}(P_{x}) = P_{x}\otimes \cos\big( \tfrac{\lambda}{2}\,P_z \big) + \cos\big(
\tfrac{\lambda}{2}\,P_z \big)\otimes P_{x} + P_{y}\otimes \sin\big( \tfrac{\lambda}{2}\,P_z \big) - \sin\big(
\tfrac{\lambda}{2}\,P_z
\big)\otimes P_{y} \ , \label{TwistedPoincareCoproduct}\\[4pt]
&\sfDelta_{\cal F}(P_{y}) = P_{y}\otimes \cos\big( \tfrac{\lambda}{2}\,P_z \big) + \cos\big(
\tfrac{\lambda}{2}\,P_z \big)\otimes P_{y} - P_{x}\otimes \sin\big( \tfrac{\lambda}{2}\,P_z \big) + \sin\big(
\tfrac{\lambda}{2}\,P_z
\big)\otimes P_{x} \ .
\end{split}
\end{align}
We can now proceed as usual to define a twisted differential geometry \cite{Aschieri2009} using (\ref{AngTwistCart}) and develop noncommutative quantum field theory as in \cite{DimitrijevicCiric:2018blz}. {The modified coproducts of momentum operators $P_x$ and $P_y$ lead to deformed momentum conservation laws.}

However, there exists a choice of local frame in which the twist (\ref{AngTwistCart}) assumes a translationally invariant Moyal-like form: on general grounds, by Frobenius' Theorem any abelian twist is locally equivalent to a Moyal twist on a dense open submanifold via a suitable change of local frame adapted to the corresponding foliation~\cite{Aschieri:2009qh}. For the angular twist \eqref{AngTwistCart}, these are the local cylindrical coordinates $(x^\mu) = (t,r,\varphi,z)$, defined for $r\in [0,\infty)$ and $\varphi\in[0,2\pi)$ by $(x,y)=(r\cos\varphi,r\sin\varphi)$, in which it becomes
\begin{equation}
\mathcal{F} = \exp\big({-\tfrac{\im \, \lambda}{2} \,(\partial_z\otimes\partial_\varphi - \partial_\varphi\otimes\partial_z)}\big) \ .
\label{AngTwist0Phi}
\end{equation} 

The twist (\ref{AngTwist0Phi}) defines the noncommutative star-product of functions $f,g\in \CCC^\infty(\FR^4)$ as
\begin{align}
f\star g :=  \sfm \circ \CF^{-1}(
f\otimes g ) = f\cdot g + \sum_{n=1}^\infty\,\bigg(\frac{\im\,\lambda}2\bigg)^n \ \sum_{k=0}^n\,\frac{(-1)^k}{k!\,(n-k)!} \, \partial_z^{n-k}\,\partial_\varphi^k f\cdot\partial_{\varphi}^{n-k}\,\partial_z^k g \ , \label{fStarg0Phi}
\end{align}
where $\sfm(f\otimes g) = f\cdot g$ is the commutative pointwise product and $\CCC^\infty(\FR^4)\subseteq C^\infty(\FR^4)$ is a suitable subspace of smooth integrable complex-valued functions on $\FR^4$ which is closed under \eqref{fStarg0Phi}. The star-product makes the space of functions $\CCC^\infty(\FR^4)$ into a $U_\CF\,\fra$-module algebra.
The exterior algebra of differential forms is deformed in a similar manner \cite{DimitrijevicCiric:2019hqq}. 

The noncommutativity of the star-product is controlled by the triangular universal $R$-matrix ${\cal R} = \CF^{-2}$ whose inverse is given by   
\begin{equation}
{\cal R}^{-1} = \sfR_{\alpha}\otimes\sfR^{\alpha} :=  \exp\big(-\im\,\lambda\,(\partial_z\otimes \partial_\varphi - \partial_\varphi\otimes
\partial_z)\big) \ , \label{RMatrix} 
\end{equation}
with implicit summation over repeated upper and lower indices understood throughout.
In particular,
\begin{align}
f\star g =  \sfR_{\alpha}(g)\star \sfR^{\alpha}(f)  \ .  \nn
\end{align}
In other words, the star-product of functions is \emph{braided commutative}.

The integration of functions over $\FR^4$ is defined in the usual way. Since the twist is abelian, the integral is cyclic with respect to the star-product (so that it defines a trace on the noncommutative algebra of functions), and the star-product is closed so that one star-product under the integral can always be traded for the pointwise product of functions, that is,
\begin{equation}
\int_{\FR^4}\,\dd^4x \ f_1 \star f_2\star \cdots \star f_n =   \int_{\FR^4}\,\dd^4x \ f_2 \star \cdots \star f_n\star f_1 = \int_{\FR^4}\,\dd^4x \ f_1\cdot (f_2 \star\cdots\star f_n) \ , \label{IntegralCyc}
\end{equation}
which follows from formal integration by parts.

\subsection{Scalar fields on \texorpdfstring{$\lambda$} {}-Minkowski space} \label{CylindricalCoordinates}%

In this paper we will analyse in detail the simplest interacting quantum field theory on $\lambda$-Minkowski space. This is the noncommutative real cubic scalar field theory whose classical action is 
\begin{align}
S_{\rm cl} = \int_{\FR^4}\,\d^4 x \ \Big( \frac{1}{2}\,\varPhi\,\big(-\square - m^2\big)\,\varPhi - \frac{g}{3!}\,\varPhi\star\varPhi\star\varPhi \Big)\ , \label{S_class}
\end{align}
 with $\varPhi\in \CCC^\infty(\FR^4)$ and $m,g\in\FR$, where $\square+m^2$ is the massive Klein-Gordon operator and the star-product is given by \eqref{fStarg0Phi}. We will also refer to this as $\varPhi_\star^3$-theory for brevity.

The usual starting point of perturbation theory is to decompose the scalar fields $\varPhi$ in a basis which diagonalises the free field theory, i.e. in a basis of eigenfunctions $\ttf_k$ of the Klein-Gordon operator
\begin{align}\label{KGEq} 
    \square\, \ttf_k = -k^2\,\ttf_k \ .
\end{align}

To solve \eqref{KGEq}, one has to choose a coordinate frame in which to represent the Klein-Gordon operator. This is usually done by using Cartesian  coordinates  $(x^\mu)=(t,x,y,z)$ where the Klein-Gordon operator is represented as
\begin{align*}
    \square = \frac{\partial^2}{\partial t^2} - \frac{\partial^2}{\partial x^2} - \frac{\partial^2}{\partial y^2} - \frac{\partial^2}{\partial z^2} \ .
\end{align*}
Separating the variables, the equation is solved by the usual plane waves $\ttf_k=\tte_k$, with
\begin{align}\label{eq:planewaves}
    \tte_k(x) := \e^{-\im\,k\,\cdot\, x} \ ,
\end{align}
where $k=(k_\mu)=(E,k_x,k_y,k_z)\in\big(\FR^{4}\big)^*$ and the dual pairing $k\,\cdot\, x = k_\mu\,x^\mu$ is with respect to the Minkowski metric. Since these states also diagonalise the momentum operators $P_{x^\mu}$, with eigenvalues $-k_\mu$, they are closely related to the translational invariance of the theory under consideration.

Here we wish to make the following elementary yet crucial point.
Unlike the Moyal twist, because the twist \eqref{AngTwist0Phi} breaks translational symmetry in the $(x,y)$-plane, it is not diagonalised by the plane wave basis. This makes the interaction of the theory defined by \eqref{S_class} quite cumbersome and difficult to handle when decomposed in this way~\cite{DimitrijevicCiric:2018blz}. However, the angular twist does preserve a cylindrical symmetry, and this suggests solving the eigenvalue problem \eqref{KGEq} for the Klein-Gordon operator instead in cylindrical coordinates, in a way that also diagonalises $\CF$ and hence exploits the symmetries of the interacting theory. This is possible to do using the solution of the three-dimensional Helmholtz equation in cylindrical coordinates, as we now explain.

We start by expressing the Klein-Gordon operator in the local coordinates $(x^\mu)=(t,r,\varphi,z)$, which reads
\begin{align*}
    \square = \frac{\partial^2}{\partial t^2} - \frac{\partial^2}{\partial z^2} - \frac{1}{r^2}\,\frac{\partial^2}{\partial \varphi^2} - \frac{1}{r}\, \frac{\partial}{\partial r} \left(r\,\frac{\partial}{\partial r}\right) \ .
\end{align*}
We use separation of variables and seek a solution $\ttf_k = \ttc_k$  to (\ref{KGEq}) in the form $$\ttc_k(x) = {\tt T}_k(t)\,{\tt Z}_k(z)\,{\tt R}_k(r)\,{\tt\Omega}_k(\varphi) \ . $$ After separating the $t$ and $z$ dependence, up to overall arbitrary constants we recover the usual plane waves
\begin{align*}
    {\tt T}_E(t) = \e^{\,\pm\, \im\, E\, t} \qquad , \qquad {\tt Z}_{k_z}(z) = \e^{\,\pm\, \im\, k_z\, z}
\end{align*}
with $E,k_z\in\FR$.

The partial differential equation \eqref{KGEq} then reduces to
\begin{equation}
\frac{{\tt\Omega}_k''(\varphi)}{{\tt\Omega}_k(\varphi)} + \frac{r}{{\tt R}_k(r)}\,\frac{\dd}{\dd r}\big(r\, {\tt R}_k'(r)\big) + r^2\,\big(E^2 - k_z^2 - k^2 \big) = 0 \ .\nn
\end{equation}
Separating the $\varphi$ dependence leads to 
\begin{equation}
{\tt\Omega}_\ell(\varphi) = \e^{\,\pm\, \im\, \ell\, \varphi} \ . \nn
\end{equation}
Since $\varphi\in[0,2\pi)$ is an angular coordinate, this is single-valued if and only if $\ell \in \mathbbm{Z}$. 

With $\varrho :=r\,\sqrt{ E^2 - k_z^2 - k^2 }$, this reduces the radial equation to the Bessel equation 
\begin{equation}\nn
\varrho^2\, {\tt R}_\ell''(\varrho) + \varrho\, {\tt R}_\ell'(\varrho) + \big(\varrho^2 - \ell^2\big)\,{\tt R}_\ell(\varrho) = 0 \ .
\end{equation}
The solutions are the Bessel functions of the first and second kind of integer order $\ell$, also known as cylindrical harmonics or cylindrical Bessel functions, which are respectively given as
\begin{equation}
{\tt R}_\ell(\varrho) = c_\ell \,J_\ell(\varrho) + d_\ell\, N_\ell(\varrho) \ , \nn
\end{equation}
where $c_\ell$ and $d_\ell$ are arbitrary constants.
Since the Bessel functions of the second kind $N_\ell(\varrho)$ have a branch point singularity at $\varrho= 0$, we set all coefficients $d_\ell=0$. Some relevant properties of the Bessel functions of the first kind $J_\ell(\varrho)$, which define entire functions on $\mathbbm{C}$ by analytic continuation, are reviewed in Appendix~\ref{app:Bessel}.

It follows that the desired diagonalising basis with cylindrical symmetry is given in terms of cylindrical harmonics as
\begin{align}\label{eq:cylharm}
    \ttc_{k}(x) = \ttc_{(E,\alpha,\ell,k_z)}(t,r,\varphi,z) := J_\ell(\alpha\, r) \ \e^{\,\im\, \ell\, \varphi} \ \e^{-\im\, E\, t} \ \e^{\,\im\, k_z\, z} \ ,
\end{align}
where the corresponding eigenvalue in \eqref{KGEq} is
\begin{align}\nn
    k^2 = E^2 -\alpha^2 - k_z^2 \ .
\end{align}
From the on-shell condition $k^2=m^2$, it follows that the variables $k = (E,\alpha,\ell,k_z)$ have the following physical interpretation: $E\in\FR$ is the energy, $k_z\in\FR$ is the linear momentum in the $z$-direction, $\alpha\in[0,\infty)$ is the magnitude of the radial momentum in the $(x,y)$-plane, and $\ell\in\RZ$ is the angular momentum about the $z$-axis --- an eigenvalue of the operator $P_\varphi=-\im\,\partial_\varphi$.

These basis functions also diagonalise the angular twist operator \eqref{AngTwist0Phi}, with eigenvalue a simple Moyal-like phase factor:
\begin{align*}
    \CF\,(\ttc_k\otimes\ttc_{k'}) = \e^{\,\frac{\ii\,\lambda}2\,(k_z\,\ell' - k_z'\,\ell) } \, (\ttc_k\otimes\ttc_{k'}) \ .
\end{align*}
This streamlines calculations enormously in the basis of cylindrical harmonics, as they are formally identical to those using the Moyal twist in the plane wave basis \eqref{eq:planewaves}: many known results of the Moyal deformation can now be simply copied over and adapted to the angular twist deformation. We will make extensive use of this simplification in the present paper.

Finally, we write the spectral decomposition of a real scalar field $\varPhi\in \CCC^\infty(\FR^4)$ in this basis as
\begin{align}\label{PhiAdB}
    \varPhi = \int_{k}\hspace{-4.5mm}\mbox{$\sum$} \ \ \phi_k \ \ttc_k \ ,
\end{align}
where $\phi_k\in\mathbbm{C}$ and we use the short-hand notation
\begin{align*}
    \int_{k}\hspace{-4.5mm}\mbox{$\sum$} \ \ := \int_{\FR\times\FR}\,\frac{{\rm d}E}{2\pi} \ \frac{{\rm d}k_z}{2\pi} \ \int_0^\infty\, \alpha \, {\rm d}\alpha  \ \frac1{2\pi}\,\sum_{\ell\in\RZ} \ .
\end{align*}
The reality condition $\varPhi^*=\varPhi$ implies $\phi_k^* = (-1)^\ell\,\phi_{k^*}$, where $k^* = (-E,\alpha,-\ell,-k_z)$.
In particular, the resummation of the plane waves \eqref{eq:planewaves} into cylindrical harmonics can be read off from~\cite[eq.~(35)]{Jackson:1972} and is given by a Fourier series transformation
\begin{align} \label{eq:planetocyl}
    \tte_{(E,k_x,k_y,k_z)} = \sum_{\ell\in\RZ} \, \ii^{\,\ell} \ \e^{-\ii\,\ell \tan^{-1}\big(\frac{k_y}{k_x}\big)} \ \ttc_{(E\,,\,\sqrt{k_x^2+k_y^2}\,,\,\ell\,,\,k_z)} \ .
\end{align}

The essential notation and the correspondence between the plane wave and cylindrical harmonic bases are summarized in Table~\ref{tab:basis-dictionary}. The notation is used in greater detail in Section~\ref{PWversusCY}.

\begin{table}[H]
\centering
\begin{tabular}{|c| | c|c|}\hline
 & \textbf{Plane waves} & \textbf{Cylindrical harmonics} \\[6pt]
\hline\hline
\textbf{Spacetime coordinates} & $(x^\mu)=(t,x,y,z)$ & $(x^\mu)=(t,r,\varphi,z)$ \\[6pt]
\hline
\textbf{Eigenfunctions} &
$\tte_k(x) = \e^{-\ii\,k\cdot x}$ &
$\ttc_k(x) = J_\ell(\alpha\, r)\, \e^{\,\ii\, \ell \varphi}\, \e^{-\ii E t}\, \e^{\,\ii\, k_z\, z}$ \\[6pt]
\hline
\textbf{Momenta (eigenvalues)} &
$k = (E, k_x, k_y, k_z)$ &
$k = (E, \alpha, \ell, k_z)$ \\[6pt]
\hline
\textbf{Relation between bases} &
\multicolumn{2}{|l|}{$\displaystyle \tte_k = \sum_{\ell\in\mathbb{Z}}\, \ii^{\,\ell}\, \e^{-\ii\, \ell\, \vartheta}\, \ttc_k \ ,  \quad \alpha = \sqrt{k_x^2+k_y^2} \ , \quad \vartheta = \tan^{-1}\!\Big(\frac{k_y}{k_x}\Big)$} \\[4pt]\hline
\end{tabular}
\caption{\small Dictionary between the plane wave basis and the cylindrical harmonic basis used throughout the paper. In both bases $E$ is the energy and $k_z$ is the axial momentum, while $\alpha$ is the magnitude of the radial momentum  in the $(x,y)$-plane, $\vartheta$ its polar angle, and $\ell$ is the angular momentum about the $z$-axis.}
\label{tab:basis-dictionary}
\end{table}

\section{Algebraic approach to Batalin-Vilkovisky quantization} \label{QFT}%

This paper is concerned with the perturbative quantum field theory based on the classical action \eqref{S_class}, from the perspective of the modern homotopy algebraic setting of the Batalin-Vilkovisky (BV) formalism and homological perturbation theory. In this section we will briefly review the basics of this algebraic approach to  BV quantization of general field theories. Detailed definitions and further details about the constructions of this section can be found in~\cite{Nguyen:2021rsa, Bogdanovic:2024jnf}. 

As discussed in~\cite{DimitrijevicCiric:2021jea,Nguyen:2021rsa,Giotopoulos:2021ieg}, the formalism makes sense in essentially any category of vector spaces endowed with a closed symmetric monoidal structure. For our purposes, there are two relevant categories of vector spaces, which in the subsequent sections will lead to two different quantum field theories that quantize the $\varPhi_\star^3$-theory defined by~\eqref{S_class}:
\begin{myitemize} 
\item The most commonly used category $\CCV$ of complex vector spaces and linear maps has trivial braiding given by simply permuting the factors of tensor products. Within this category, one reproduces the standard path integral approach to quantum field theory, and in particular the standard quantization of noncommutative field theories~\cite{Giotopoulos:2021ieg}, albeit in a purely algebraic fashion without reference to functional integration. We will refer to this as the `standard BV formalism'. 

\item The representation category $\CCR$ of modules and equivariant maps for the twisted Hopf algebra $U_\CF\,\fra$ associated to the angular twist operator \eqref{AngTwistCart} has non-trivial braiding defined in terms of the $R$-matrix $\CR$. Working in this category leads to braided quantum field theory~\cite{DimitrijevicCiric:2021jea,Nguyen:2021rsa,Giotopoulos:2021ieg}, which is entirely based on the Drinfel'd twist formalism and has no path integral description. We will refer to this as the `braided BV formalism'. It is important to stress that, in this latter approach, the constraints of $U_\CF\,\fra$-equivariance must be imposed on all objects and maps in order to ensure that all standard homotopical constructions (e.g. the Homological Perturbation Lemma) carry over to this generalised setting; indeed, this is one of the motivations for introducing the basis of cylindrical harmonics in Section~\ref{CylindricalCoordinates}.
\end{myitemize}
In this section we work in the former category $\CCV$ in order to most clearly illustrate the formalism. 

Let $(V,\{\mu_n\}_{n\geqslant 1})$ be the  $L_\infty$-algebra in $\CCV$ which organises the dynamics of a classical field theory containing  polynomial interactions, with or without gauge symmetries. The graded vector space $$V=\bigoplus_{k\in\RZ}\,V^k$$ contains all the fields of the theory as well as the corresponding antifields. We write $\CA$ for a generic field of $V$ and $\CA^+$ for its antifield; if $\CA\in V^k$ for $k\leqslant1$, then $\CA^+\in V^{3-k}$. We denote by $|\CA|$ the degree of a homogeneous element $\CA\in V$; if $\CA\in V^k$, then $|\CA| = k$. The multilinear graded antisymmetric $n$-brackets $$\mu_n:V^{\otimes n}\longrightarrow V[n-2]$$ obey a sequence of homotopy Jacobi identities, where $V[l]$ for $l\in\RZ$ is the vector space $V$ with its grading shifted according to $(V[l])^k = V^{k+l}$. 

In the case of a Lagrangian field theory, its $L_\infty$-algebra is further equipped with a cyclic graded symmetric non-degenerate pairing $$\langle-,-\rangle : V \otimes V \longrightarrow \FC[-3]$$ between fields and their antifields, usually chosen to be an integration over spacetime or a trace. Then the classical action of the theory is captured by the Maurer-Cartan functional
\begin{align}\nn
    S_\MC = \frac12\,\langle A,\mu_1(A)\rangle + \sum_{n\geqslant2} \, \frac{(-1)^{n\choose 2}}{(n+1)!} \, \langle A,\mu_n(A^{\otimes n})\rangle
\end{align}
on the physical fields $A\in V^1$. Its variational equation is the Maurer-Cartan equation of the $L_\infty$-algebra, which captures the classical equations of motion of the  field theory.

The free field theory is determined by the lowest bracket $\mu_1:V\longrightarrow V[-1]$, which is a differential making $(V,\mu_1)$ into a cochain complex. The cohomology $H^\bullet(V)$ of this complex is the space of classical vacua of the BV theory; it is also a cochain complex with the trivial differential. The complex $(V,\mu_1)$ is quasi-isomorphic to $(H^\bullet(V),0)$. For this, we introduce a homotopy equivalence through a degree zero inclusion map $\mathsf{i}\,$, a degree zero projection map $\sfp$ and a contracting homotopy $\sfh$ of degree $-1$, which together satisfy the conditions that define a (strong) retract of $(V,\mu_1)$ onto $(H^\bullet(V),0)$:
\begin{equation}\label{SDR1}
\begin{tikzcd}
\arrow[out=120,in=60,loop,looseness=3,"\sfh"] (V,\mu_1) \ar[r,shift right=1ex,swap,"\sfp"] & \ar[l,shift right=1ex,swap,"\mathsf{i}"] (H^\bullet(V),0) 
\end{tikzcd} \ .
\end{equation}
The map $\sfh$ is a cochain homotopy equivalence between $\id_V$ and $\mathsf{i}\circ\sfp$, that is,
\begin{align*}
    \mathsf{i}\circ\sfp = \id_V+\mu_1\circ\sfh+\sfh\circ\mu_1 \ ,
\end{align*}
which is determined by the propagator of the free theory.

Quantization and interactions among fields can be introduced once we extend these data to the space of functionals $$\Sym(V[1]^*)=\bigoplus_{n=0}^\infty \,(V[1]^*)^{\odot n}$$ on $V[1]$, which are the classical polynomial observables of the BV theory. This is the unital graded commutative algebra generated by symmetric products of coordinate functionals on $V[1]$, with product denoted
\begin{align}\nn
    F\odot F' = (-1)^{|F|\,|F'|} \ F'\odot F \ ,
\end{align}
and $(V[1]^*)^{\odot\, 0}:=\FC$ playing the role of the constant functionals. Using the non-degenerate pairing $\langle-,-\rangle$ we can identify $V[1]^*= V[2]$: the duality pairing between $V[2]$ and $V[1]$ is given by
\begin{align*}
    \langle-,-\rangle[3]:V[2]\otimes V[1] = (V\otimes V)[3] \longrightarrow \FC[-3][3] = \FC \ .
\end{align*}

Then $\big(\Sym(V[2]),\mu_1\big)$ is the derived space of observables of the free BV theory, where the differential $\mu_1$ is extended as a graded derivation of degree zero to the entire algebra $\Sym (V[2])$.
In this way we obtain a new homotopy equivalence of cochain complexes
\begin{equation}\label{SDR2}
\begin{tikzcd}
\arrow[out=160,in=20,loop,looseness=2,"\sH"] \big(\Sym(V[2]),\mu_1\big) \ar[r,shift right=1ex,swap,"\sP"] & \ar[l,shift right=1ex,swap,"\sI"] \big(\Sym (H^\bullet(V[2])),0\big) 
\end{tikzcd} \ .
\end{equation}
 The cochain maps $\sP$, $\sI$ and $\sH$ are the ``thickened'' extensions of the corresponding cochain maps in~(\ref{SDR1}). 

The data are further supplemented with the BV antibracket which is the graded shifted Poisson bracket
\begin{equation}\label{antibracket}
    \{-,-\}:\Sym (V[2])\otimes \Sym (V[2])\longrightarrow \Sym (V[2])[1]
\end{equation}
that is compatible with the differential $\mu_1$.
It is defined by its non-zero action on generators from $V$ as
\begin{equation}\nn
    \{\CA,\CA^+\} := \langle \CA,\CA^+\rangle \ ,
\end{equation}
and extended as a graded derivation on $\Sym(V[2])$ in each of its slots.

Given the homotopy equivalence (\ref{SDR2}), we now make a small perturbation of the free BV differential $\mu_1$ to the differential $$Q_{\mbf\delta} = \mu_1 + {\mbf\delta} \ . $$ Then the Homological Perturbation Lemma constructs perturbed cochain maps $\sP_{\mbf\delta}$, $\sI_{\mbf\delta}$ and $\sH_{\mbf\delta}$ that define another homotopy equivalence of complexes. 
In the category of vector spaces $\CCV$, the perturbation of the projection $\sP_{{\mbf\delta}}:\Sym(V[2])\longrightarrow \Sym (H^\bullet(V[2]))$ implements the path integral via homotopy transfer~\cite{Doubek:2017naz}.

The (smeared) physical $n$-point correlation functions of the quantum field theory are then defined as~\cite{Nguyen:2021rsa}
\begin{align}\label{eq:BQFTnpoint}
\begin{split}
G_n(\ttf_1,\dots,\ttf_n) := \sP_{{\mbf\delta}}(\ttf_1\odot\cdots\odot\ttf_n)  = \sum_{m=1}^\infty \, \sP\,\big(({{\mbf\delta}} \,\sH)^m\,(\ttf_1\odot\cdots\odot\ttf_n)\big) \ ,
\end{split}
\end{align}
where the external states $\ttf_a\in V^2$ are test functions corresponding to insertions of the physical fields $A_a\in V^1$.  We are interested in two important perturbations ${\mbf\delta}$: 
\begin{myitemize}
\item The free quantum field theory is defined by ${\mbf\delta}_\hbar = -\ii\,\hbar\,\BVL$ with the BV Laplacian $$\BVL:\Sym( V[2])\longrightarrow \Sym (V[2])[1] \ .$$
The properties of the BV Laplacian follow from the fact that the perturbed differential is again a differential, $Q_{\mbf\delta_\hbar}^2 = (\mu_1-\ii\,\hbar\,\BVL)^2 = 0$, and its definition as a graded derivation on $\Sym (V[2])$ up to the antibracket (\ref{antibracket}). Explicitly, it is defined on generators by
\begin{align}\nn
    \BVL(1)=0 \quad , \quad \BVL(\CA) = 0 = \BVL(\CA^+) \quad , \quad \BVL(\CA\odot\CA^+) = \{\CA,\CA^+\} = \langle\CA,\CA^+\rangle \ ,
\end{align}
and extended to arbitrary elements $F,F'\in\Sym(V[2])$ as
\begin{align}\nn
    \BVL(F\odot F') = \BVL(F)\odot F' + (-1)^{\vert F\vert}\, 
F\odot\BVL(F') +  \{F,F'\} \ .
\end{align}
The BV Laplacian implements Wick's Theorem in correlation functions.

\item The interacting field theory is defined by ${\mbf\delta}_{\rm int} = - \{\CS _{\rm int},-\}$, where $\CS_{\rm int}$ is the interacting part of the BV master action, defined as~\cite{Jurco:2018sby}
\begin{equation}
\CS_{\rm int} = \sum_{n\geqslant2} \, \frac{(-1)^{n\choose 2}}{(n+1)!} \, \langle\xi,\mu_n^{\rm ext}(\xi^{\otimes n})\rangle^{\rm ext} \ \in \ \Sym(V[2]) \ .\label{S_celo_BV}
\end{equation}
Here the contracted coordinate functions $\xi$ are degree one elements of $\Sym(V[2])\otimes V$ defined by $\xi = \ttv^I\otimes \ttv_I$, where $\ttv_I$ is a basis for $V$ with dual basis $\ttv^I = \ttv_I^+$ with respect to the cyclic pairing. The superscripts ${}^{\rm ext}$ denote the extension of the cyclic $L_\infty$-structure on the (shifted) space of BV fields $V$ to the tensor product $\Sym(V[2])\otimes V$ with the graded commutative algebra of classical observables~$\Sym(V[2])$. 
\end{myitemize}

With the sum $\mbf\delta = \mbf\delta_\hbar + \mbf\delta_{\rm int}$ of these two perturbations, which defines the interacting quantum BV differential $Q_{\mbf\delta}$,  the expression \eqref{eq:BQFTnpoint} is a formal power series in Planck's constant $\hbar$ and in the coupling constants appearing in $\CS_{\rm int}$ with coefficients valued in $\Sym (H^\bullet(V[2]))$. That is, the quantum correlator \eqref{eq:BQFTnpoint} is a functional on the space of classical vacua. Upon evaluation on a particular vacuum state, this represents the perturbative expansion of the quantum field theory about that vacuum.

The relation between the standard and braided BV formalisms discussed at the beginning of this section can now be clarified through the lens of $A_\infty$-algebras. An $A_\infty$-structure has no symmetries and so is agnostic to the braiding of the monoidal category. Given an $A_\infty$-algebra in the category $\CCR$, braided graded antisymmetrization of its brackets results in a braided $L_\infty$-structure. On the other hand, this same $A_\infty$-structure defines an $A_\infty$-algebra in the category $\CCV$ under the forgetful functor $\CCR\longrightarrow\CCV$ which forgets the  $U_\CF\,\fra$-module structures, and standard graded antisymmetrization of its brackets now leads to an ordinary $L_\infty$-structure. In fact, all elements of homological perturbation theory apply to $A_\infty$-algebras, and (braided) quantum field theory can be developed equally well in the language of $A_\infty$-structures by replacing the (braided) symmetric algebra $\Sym(V[2])$ above with the free unital tensor algebra ${\rm T}(V[2]) = \bigoplus_{n\geqslant0}\,(V[2])^{\otimes n}$ over $V[2]$. 

\section{Braided Batalin-Vilkovisky formalism for $\mbf{\varPhi_\star^3}$-theory} \label{BraidedQFT}%

The homotopy algebra underlying the noncommutative cubic scalar field theory defined by (\ref{S_class}) is most straightforwardly constructed by applying the Drinfel'd twist deformation formalism, with the angular twist operator \eqref{AngTwist0Phi}, directly to the $L_\infty$-algebra which organises the corresponding commutative theory; the latter is in fact a differential graded Lie algebra (in the category of $L_\infty$-algebras). In this way we obtain the braided $L_\infty$-algebra which completely determines the classical $\varPhi_\star^3$-theory (in the category of braided $L_\infty$-algebras). From this perspective, it is more natural to start with the quantization of $\varPhi_\star^3$-theory as a braided quantum field theory. In this section we give an elaborate account of this.

\subsection{$L_\infty$-structure}

Consider the twisted $L_\infty$-structure $(V, \mu_1^\CR,\mu_2^\CR)$ with underlying graded vector space $$V=V^1\oplus V^2$$ consisting of physical fields $\varPhi\in V^1=\CCC^\infty(\FR^4)$ and their antifields $\varPhi^+\in V^2=\CCC^\infty(\FR^4)$. The non-vanishing brackets are given by
\begin{align} \label{braidedell}
\begin{split}
\mu^\CR_1 (\varPhi) &= \big(-\square -m^2\big)\, \varPhi \ , \\[4pt]
\mu^\CR_2 (\varPhi_1, \varPhi_2) &= g\ \varPhi_1\star\varPhi_2 = \mu^\CR _2 \big(\sfR_\alpha(\varPhi_2), \sfR^\alpha(\varPhi_1)\big) \ . 
\end{split}
\end{align}
Since the binary bracket $\mu_2^\RR$ is braided symmetric, this defines a braided $L_\infty$-algebra.

This braided $L_\infty$-algebra is strictly cyclic, with cyclic pairing
\begin{equation}
\langle \varPhi, \varPhi ^+\rangle_\CR  := \int_{\FR^4}\,\d^4 x \ \varPhi \star \varPhi^+ = \int_{\FR^4}\,\d^4 x \ \varPhi \cdot \varPhi^+  \ .
\label{CyclPairing}
\end{equation}
The classical action (\ref{S_class}) is then reproduced by the corresponding Maurer-Cartan functional
\begin{equation}
    S_{\rm cl} = \tfrac{1}{2!} \, \langle \varPhi, \mu^\CR_1(\varPhi)\rangle_\CR - \tfrac{1}{3!} \, \langle\varPhi, \mu^\CR_2(\varPhi, \varPhi)\rangle_\CR \ . \nn 
\end{equation}

The cohomology $H^\bullet(V)$ of the underlying cochain complex $(V,\mu_1^\CR)$ describes the classical vacua of the free $\varPhi_\star^3$-theory. It is itself a cochain complex concentrated in degrees one and two, given by the solution space $H^1(V)=\ker(\square+m^2)$ of the massive Klein-Gordon equation and the space $H^2(V)={\rm coker}(\square+m^2)$, with the trivial differential~$0$.

To set up a retract \eqref{SDR1} which determines a quasi-isomorphism between the complexes $(V,\mu_1^\CR)$ and $(H^\bullet(V),0)$ in the representation category of $U_\CF\,\fra$-modules $\CCR$, we need to find a homotopy $\sfh: V^2\longrightarrow V^1$ which is a $U_\CF\,\fra$-invariant map~\cite{Nguyen:2021rsa}. In contrast to the case of the usual Moyal twist, where full translational symmetry is preserved by the noncommutative deformation, here the projection map $\sfp: V\longrightarrow H^\bullet(V)$ must be equivariant under the action of translations and  rotations with respect to the $z$-direction. 

The symmetries respected by the angular twist deformation uniquely select the basis vectors of $V$ to be the cylindrical harmonics of Section~\ref{CylindricalCoordinates}, as the usual plane wave basis \eqref{eq:planewaves} in Cartesian coordinates is incompatible with the $U_\CF\,\fra$-equivariance imposed by the twist \eqref{AngTwist0Phi}. In the notation of \eqref{eq:cylharm}, we take
\begin{align}
\ttc_k (x) :\!&= \ttc_{(E, \alpha, \ell, k_z)} (t, r, \varphi, z) \  \in \ V^1 \ ,\label{BasisFields}\\[4pt]
\ttc^k (x) :\!&= \ttc_{k^*}(x) =  \ttc_{(-E, \alpha, -\ell, -k_z)}(t, r, \varphi, z)  \ \in \ V^2 \ . \label{BasisAntiFields}
\end{align}
Since the cylindrical Bessel functions satisfy the orthogonality relation \eqref{2Bessel}, the duality between these basis vectors in the spaces of fields and antifields follows as
\begin{align}\label{eq:cylpairing}
\begin{split}
\langle \ttc_{k_1} \, , \ttc^{k_2} \rangle_\CR &= \int_{\FR\times\FR}\, \d t \ \d z \ \int_0^\infty\, r \, \d r \ \int_0^{2\pi}\,\d \varphi \ J_{\ell_{1}}(\alpha_{1}\, r) \ \e^{\,\im\, \ell_{1}\, \varphi} \ \e^{-\im\, E_1\, t} \ \e^{\,\im\, k_{1z}\, z} \\[-8pt]
& \hspace{6cm} \times \, J_{\ell_{2}}(\alpha_{2}\, r) \ \e^{- \im\, \ell_{2}\, \varphi} \ \e^{\,\im\, E_2\, t} \ \e^{- \im\, k_{2z}\, z} \\[4pt]
& = (2\pi)^3\ \delta(E_1 - E_2)\ \delta(k_{1z} - k_{2z}) \ \delta_{\ell_{1} , \ell_{2}}\ \frac{\delta(\alpha_{1} - \alpha_{2})}{\alpha_{1}} \ .
\end{split}
\end{align}

Consider now the Poincar\'e-invariant Feynman propagator
\begin{align}
\sgreen = -\frac1{\square+m^2} \quad , \quad \widetilde \sgreen(k) = \frac1{k^2-m^2-\ii\,0^+} \ ,\label{TildeG}
\end{align}
where $\widetilde \sgreen(k)$ denotes the eigenvalues of the Green operator $\sgreen$ when acting on the basis vectors \eqref{BasisAntiFields}, with $k^2 =  E^2 -\alpha^2 - k_z^2$. The Green operator satisfies
\begin{align*}
\sgreen\circ\mu_1^\CR = -\sgreen\circ\big(\square+m^2\big) = \id_{V^1} \qquad , \qquad \mu_1^\CR\circ \sgreen = -\big(\square+m^2\big)\circ \sgreen = \id_{V^2} \ .
\end{align*}

The retract conditions of \eqref{SDR1} can now be solved. For example, consider  the projection $\sfp=0$ onto the trivial vacuum state $\varPhi=\varPhi^+=0$. Then the cochain homotopy equivalence
$\sfh:V^2\longrightarrow V^1$ between $\id_V$ and $\mathsf{i}\circ\sfp=0$ is given by $\sfh = -\sgreen$. Explicitly,
\begin{align}\label{eq:htwo}
\sfh(\varPhi^+)_k = -\frac1{k^2 - m^2 -\ii\,0^+}\, \phi^+_k =  -\widetilde \sgreen(k)\,\phi^+_k \ ,
\end{align}
for $\varPhi^+\in V^2$. In the following we abbreviate $$m_+^2:= m^2+\ii\,0^+$$ inside loop integrals.

\subsection{Quantum correlation functions}

Since the  $L_\infty$-algebra organising the classical theory is a braided $L_\infty$-algebra, i.e. an $L_\infty$-algebra in the twisted Hopf representation category $\CCR$, which carries a non-trivial (but symmetric) braiding, all maps defined in Section~\ref{QFT} have to be consistently deformed by the twist (\ref{AngTwist0Phi}) in order to define morphisms in this category. In particular, the algebra of classical observables on $V[1]$ is now the unital braided symmetric algebra $$\Sym_\RR (V[2]) = \bigoplus_{n=0}^\infty \,(V[2])^{\odot_\CR n} \ , $$ defined via the twisted symmetric product $\odot_\CR$ which is braided graded commutative:
\begin{align}\label{SymR}
F\odot_\CR F' = (-1)^{|F|\,|F'|} \ \sfR_\alpha(F')\odot_\CR\sfR^\alpha(F) \ .
\end{align}

Following the prescription of Section~\ref{QFT}, the cochain maps $\sfp$ and $\sfh$ are extended to cochain maps $\sP$ and $\sH$ on the noncommutative algebra of classical observables $\Sym_\RR (V[2])$. We are interested in the computation of vacuum correlation functions, which requires restricting to the trivial projection map $\sfp=0$~\cite{Okawa:2022sjf,DimitrijevicCiric:2023hua}. 

To introduce interactions and quantize the theory, we first  extend the  differential $\mu_1^\CR$ as a strict graded derivation of degree zero to the entire algebra $\Sym_\RR (V[2])$. We then perturb this free BV differential to the interacting quantum BV differential
\begin{align*}
Q_{\mbf\delta^\CR} = \mu_1^\CR + \mbf\delta^\CR
\end{align*}
on $\Sym_\RR(V[2])$, with 
\begin{align*}
\mbf\delta^\CR = -\ii\,\hbar\,\BVL - \{\CS^\CR_{\rm int},-\}_\CR \ .
\end{align*}
The cochain map $\{-,-\}_\CR$ is the angular twist deformation of the classical antibracket $\{-,-\}$; in particular, it is now braided graded symmetric and defines a braided graded derivation on $\Sym_\CR(V[2])$ in each of its slots. The formal definition of the BV Laplacian $\BVL$ on $\Sym_\CR(V[2])$ is unchanged from Section~\ref{QFT}, but it now implements the Braided Wick Theorem in correlation functions; further details can be found in~\cite{Nguyen:2021rsa, Bogdanovic:2024jnf}. 

The correlation functions of the (interacting) braided quantum field theory are written in a momentum space representation analogous to \eqref{eq:BQFTnpoint} as
\begin{equation}\label{eq:intcorrelationfnMomentum}
\widetilde C_n^\CR(p_1,\dots,p_n) = \sum_{m=1}^\infty \, \sP\,\big((-\ii\,\hbar\,\BVL\,\sH - \{\CS^\CR _{\rm int},-\}_\CR\,\sH)^m\, (\ttc^{p_1}\odot_\CR\cdots\odot_\CR\ttc^{p_n})\big) \ ,
\end{equation}
where the external states are the antifields $\ttc^{p_a}$ defined in (\ref{BasisAntiFields}) and all momenta $p_a=(E_a,\alpha_a,\ell_a,p_{za})$ are treated as incoming in the propagators.
Since $\sfp=0$, only $\sP(1) = 1$ is non-zero, and hence this is a formal power series expansion in Planck's constant $\hbar$ and the coupling constant $g$. 

The definition of the quantum correlator \eqref{eq:intcorrelationfnMomentum} has the following representation theoretic meaning. Recall that the fields and antifields are naturally superpositions of $U_\CF\,\fra$-invariant cylindrical harmonics.
By construction, the $n$-point function \eqref{eq:intcorrelationfnMomentum} is determined by the perturbed projection map $$\sP_{\mbf\delta^\RR}:(V[2])^{\odot_\CR n}\longrightarrow \FC$$ from the $n$-th irreducible (braided) symmetric representation of the two-dimensional abelian Lie algebra $\fra$, with weight $(p_{1z}+\cdots+p_{nz},\ell_{1}+\cdots+\ell_{n})$, to the trivial representation $\FC$. This map is $U_\CF\,\fra$-equivariant if and only if these weight spaces are themselves isomorphic to the trivial representation, that is,
\begin{align}\nn
  p_{1z}+\cdots+p_{nz}=0 \qquad , \qquad \ell_1+\cdots+\ell_n = 0 \ .  
\end{align}

Thus the braided quantum field theory in the cylindrical harmonic basis naturally imposes the kinematical conservation of axial and angular momenta in correlation functions. This is appropriate here since the theory does not exhibit the full translational invariance, as translational symmetry in the $(x,y)$-plane is broken, but rather has a cylindrical symmetry. This would not be true in the standard basis of plane waves \eqref{eq:planetocyl}, which do not span irreducible representations of $\fra$ and instead lead to a deformed momentum conservation law in correlation functions. We will return to this point in Section~\ref{sub:planewave}.

\subsection{Interacting Batalin-Vilkovisky master action} \label{sub:intBVbraided}

The interaction functional $\CS _{\rm int}^{\CR}\in \Sym_\RR(V[2])$ for braided  $\varPhi_\star^3$-theory is defined by
\begin{equation}\nn
\CS_{\rm int}^{\RR} := -\tfrac{1}{3!}\,\langle \xi \,,\,\mu^{\RR \, {\rm ext}}_2(\xi,\xi)\rangle^{\rm ext}_\RR \ .    
\end{equation}
In the notation of \eqref{PhiAdB}, the contracted coordinate functions are given by
\begin{align*}
\xi = \int_{k}\hspace{-4.5mm}\mbox{$\sum$} \ \ \big(\ttc_k\otimes \ttc^k + \ttc^k\otimes \ttc_k\big) \ \in \ \Sym_\RR (V[2])\otimes V \ ,
\end{align*}
with the basis functions (\ref{BasisFields}) and (\ref{BasisAntiFields}). Crucially, $\xi$ is a $U_\CF\,\fra$-invariant element of $\Sym_\RR (V[2])\otimes V$.

Repeating the same calculations as in~\cite{Bogdanovic:2024jnf}, after applying all the extended operations using their definitions we get
\begin{equation} \label{Sint_expl}
\CS_{\rm int}^{\RR} = -\frac{g}{3!} \ \int_{k_1}\hspace{-6mm}\mbox{$\sum$} \ \ \int_{k_2}\hspace{-6mm}\mbox{$\sum$} \ \ \int_{k_3}\hspace{-6mm}\mbox{$\sum$} \ \ 
\ttc^{k_1}\odot_\RR \sfR_{\beta}(\ttc^{k_2}) \odot_\RR \sfR_{\rho}\,\sfR_{\alpha}(\ttc^{k_3}) \ \langle \sfR^{\beta}\,\sfR^{\rho}(\ttc_{k_1}) \,,\, \sfR^{\alpha}(\ttc_{k_2}) \star \ttc_{k_3}\rangle_\RR \ .
\end{equation}
Evaluating the actions of the $R$-matrices leads to a formally identical result as for the case of the Moyal twist~\cite{Bogdanovic:2024jnf} in terms of phase factors:
\begin{eqnarray*}
\CS_{\rm int}^{\RR} = -\frac{g}{3!} \ \int_{k_1}\hspace{-6mm}\mbox{$\sum$} \ \ \int_{k_2}\hspace{-6mm}\mbox{$\sum$} \ \ \int_{k_3}\hspace{-6mm}\mbox{$\sum$} \ \  \e^{\, \ii \, \lambda \, \sum\limits_{a<b}\, (k_{az}\, \ell_{b} - k_{bz} \, \ell_{a})} \ \ttc^{k_1}\odot_\RR \ttc^{k_2} \odot_\RR \ttc^{k_3} \ \langle \ttc_{k_1} \,,\, \ttc_{k_2} \star \ttc_{k_3}\rangle_\RR \ .
\end{eqnarray*}

Explicit calculation of the cyclic pairing gives 
\begin{align}\nn
\begin{split}
\langle \ttc_{k_1} \,,\, \ttc_{k_2} \star \ttc_{k_3}\rangle_\RR &= \e^{-\frac{\ii\,\lambda}{2}\, (k_{1z} \, \ell_{2} - \ell_{1} \, k_{2z})} \  \e^{-\frac{\ii\,\lambda}{2} \,((k_{1z} + k_{2z}) \, \ell_{3} - (\ell_{1} + \ell_{2}) \, k_{3z})} \\
& \quad \, \times \int_{\FR\times\FR}\, \d t \ \d z \ \int_0^\infty\, r \, \d r \ \int_0^{2\pi} \, \d \varphi \  J_{\ell_{1}}(\alpha_{1}\, r) \ J_{\ell_{2}}(\alpha_{2}\, r) \ J_{\ell_{3}}(\alpha_{3}\, r) \\
& \hspace{5cm} \times \e^{\,\ii \, (\ell_{1}+\ell_2+\ell_3)\, \varphi} \ \e^{-\ii\,( E_1 +E_2+E_3)\,t} \ \e^{\,\ii \,( k_{1z} + k_{2z} + k_{3z})\, z} \\[4pt]
&  = \e^{ -\frac{\ii\,\lambda}{2}\, \sum\limits_{a<b}\, (k_{az}\, \ell_{b} - k_{bz}\, \ell_{a})} \ (2\pi)^3 \ \delta(E_1+E_2+E_3) \ \delta(k_{1z}+k_{2z}+k_{3z}) \\
& \hspace{7cm} \times \delta_{\ell_{1}+\ell_{2}+\ell_{3}, 0} \ \ttF_{\ell_{1}, \ell_{2}, \ell_{3}} (\alpha_{1}, \alpha_{2}, \alpha_{3}) \ . 
\end{split}
\end{align}
Here the function $\ttF_{\ell_{1}, \ell_{2}, \ell_{3}} (\alpha_{1}, \alpha_{2}, \alpha_{3})$ is given as an integral of the product of three cylindrical Bessel functions 
\begin{equation} \label{Besself}
\ttF_{\ell_{1}, \ell_{2}, \ell_{3}} (\alpha_{1}, \alpha_{2}, \alpha_{3}) :=  \int_0^\infty\, r \, \d r \ J_{\ell_{1}}(\alpha_{1}\, r) \ J_{\ell_{2}}(\alpha_{2}\, r) \ J_{\ell_{3}}(\alpha_{3}\, r) \ .
\end{equation}
With the constraint $\ell_{1} + \ell_{2} + \ell_{3} = 0$, which is satisfied here due to angular momentum conservation, this integral can be computed explicitly \cite{Bailey:1936, Gervois:1984, Jackson:1972}. The result is non-zero only when certain conservation laws for the radial momenta are satisfied, as discussed in Appendix~\ref{app:Bessel} where explicit formulas may also be found.

Finally, the interacting BV master action can be written as
\begin{align}
\begin{split}
\CS_{\rm int}^{\RR} &= \int_{k_1}\hspace{-6mm}\mbox{$\sum$} \ \ \int_{k_2}\hspace{-6mm}\mbox{$\sum$} \ \ \int_{k_3}\hspace{-6mm}\mbox{$\sum$} \ \  V_\RR(k_1, k_2, k_3) \ \ttc^{k_1}\odot_\RR \ttc^{k_2} \odot_\RR \ttc^{k_3}  \ ,  \label{SintFinal} 
\end{split}
\end{align}
where
\begin{align}
\begin{split}
V_\RR(k_1, k_2, k_3) &= -\frac{g}{3!} \  \e^{\,\frac{\ii\, \lambda}{2}\, \sum\limits_{a<b}\, (k_{az}\, \ell_{b} - k_{bz}\, \ell_{a})} \ (2\pi)^3 \ \delta(E_1+E_2+E_3) \ \delta(k_{1z}+k_{2z}+k_{3z})
 \\
&\hspace{7cm} \times  \delta_{\ell_{1} + \ell_{2} + \ell_{3}, 0} \ \ttF_{\ell_{1}, \ell_{2}, \ell_{3}} (\alpha_{1}, \alpha_{2}, \alpha_{3})  \ . \label{vertex2}
\end{split}
\end{align}

The vertex \eqref{vertex2} consists of a phase factor, Dirac distributions that are invariant under any permutation of the corresponding momenta \{$E_a, k_{az}, \ell_{a}$\}, and the integral of cylindrical Bessel functions $\ttF_{\ell_{1}, \ell_{2}, \ell_{3}} (\alpha_{1}, \alpha_{2}, \alpha_{3})$ which by its definition is trivially invariant under any permutation of the pairs of momenta ($\alpha_{a}, \ell_{a}$):
\begin{equation} \label{eq:Fellalphasym}
\ttF_{\ell_{1}, \ell_{2}, \ell_{3}} (\alpha_{1}, \alpha_{2}, \alpha_{3}) = \ttF_{\ell_{2}, \ell_{1}, \ell_{3}} (\alpha_{2}, \alpha_{1}, \alpha_{3}) = \ttF_{\ell_{1}, \ell_{3}, \ell_{2}} (\alpha_{1}, \alpha_{3}, \alpha_{2}) \ .
\end{equation}
From these elementary facts it is easy to derive the braided symmetries
\begin{align}\label{VPermAdB}
V_\RR(k_1, k_2, k_3) = \e^{\,\ii\, \lambda\, (k_{1z}\, \ell_{2} - k_{2z}\, \ell_{1})} \ V_\RR(k_2, k_1, k_3) = \e^{\,\ii\, \lambda\, (k_{2z}\, \ell_{3} - k_{3z}\, \ell_{2})} \ V_\RR(k_1,k_3,k_2) \ ,
\end{align}
and, after applying the delta-functions $\delta(k_{2z}+k_{1z}+k_{3z})$ and $\delta_{\ell_{2} + \ell_{1} + \ell_{3}, 0}$, the cyclic symmetry
\begin{align}
V_\RR(k_1, k_2, k_3) = V_\RR(k_2, k_3, k_1)\ .
\label{VCyclicPermAdB}
\end{align}

Thus the interaction vertex \eqref{vertex2} is completely analogous to the corresponding Moyal-deformed interaction vertex computed in the usual basis of plane waves~\cite{Bogdanovic:2024jnf}. As a particular consequence, the perturbative expansions of correlation functions may involve both planar and non-planar contributions.

\subsection{Lowest multiplicity tree-level  functions}%

In order to illustrate the calculations in the cylindrical harmonic basis and to gain some physical intuition behind this choice of basis, we first compute the non-trivial tree-level contributions to the lowest multiplicity correlation functions \eqref{eq:intcorrelationfnMomentum}.

\subsubsection*{Two-point function}

The tree-level contribution to the two-point function $\widetilde C_2^\RR(p_1, p_2)$ is just the free propagator
\begin{equation*}
\widetilde C_2^\RR(p_1, p_2)^{\swzero} = -\ii\,\hbar\,\BVL\, \sH \, (\ttc^{p_1} \odot_\RR \ttc^{p_2}) = -\tfrac{1}{2}\,\ii\,\hbar\,\BVL \,\big[\big(\sfh(\ttc^{p_1}) \odot_\RR \ttc^{p_2}\big) + \big(\ttc^{p_1} \odot_\RR \sfh(\ttc^{p_2})\big)\big] \ .
\end{equation*}
The first term is given by
\begin{align*}
\BVL \big(\sfh(\ttc^{p_1}) \odot_\RR \ttc^{p_2}\big) &= - (2\pi)^3 \ \delta(E_1 + E_2) \ \delta(p_{1z} + p_{2z}) \ \delta_{\ell_{1} + \ell_{2}, 0} \\
& \hspace{4cm} \times \widetilde\sgreen(p_1) \  \int_0^\infty\, r \, \d r  \ J_{\ell_{1}}(\alpha_{1}\, r) \ J_{\ell_{2}}(\alpha_{2}\, r) \ .
\end{align*}
Using the orthogonality relation \eqref{2Bessel} for cylindrical Bessel functions we obtain
\begin{align*}
\BVL \big(\sfh(\ttc^{p_1}) \odot_\RR \ttc^{p_2}\big) &= - (2\pi)^3\, (-1)^{\ell_1}  \ \delta(E_1 + E_2) \ \delta(p_{1z} + p_{2z}) \ \delta_{\ell_{1} + \ell_{2},0} \ \frac{\delta(\alpha_{1} - \alpha_{2})}{\alpha_{1}} \ {\widetilde{\sgreen}}(p_1) \ .
\end{align*}

The contribution from the second term is identical and adding the two contributions results in
\begin{equation}
\widetilde C_2^\RR(p_1, p_2)^{\swzero} = \ii\,\hbar\,(2\pi)^3\, (-1)^{\ell_{1}} \ \delta(E_1 + E_2) \ \delta(p_{1z} + p_{2z}) \ \delta_{\ell_{1} + \ell_{2}, 0} \ \frac{\delta(\alpha_{1} - \alpha_{2})}{\alpha_{1}} \ {\widetilde{\sgreen}}(p_1) \ . \label{G2TreeLevel}
 \end{equation}
 In addition to the conservation of axial and angular momenta, the energy and the magnitude of the radial momentum are also conserved, as anticipated.

\subsubsection*{Three-point function}

The tree-level contribution to the three-point function $\widetilde C_3^\RR(p_1, p_2, p_3)$ can be computed starting from
\begin{equation}
\widetilde C_3^\RR(p_1, p_2,p_3)^{\swzero} =  -(\ii\,\hbar\,\BVL\,\sH)^2 \,\big\{ \CS _{\rm int}^\RR,\sH\, (\ttc^{p_1}\odot_\RR \ttc^{p_2}\odot_\RR\ttc^{p_3})\big\}_\RR \ . \label{G3Tree}
\end{equation}
Repeating the analogous calculations of~\cite{Bogdanovic:2024jnf}, the one-particle irreducible contributions obtained via the Braided Wick Theorem are of the form
\begin{align}
\begin{split}
\widetilde C_3^\RR(p_1, p_2,p_3)^{\swzero}_{\1PI} &= (\ii\,\hbar)^2 \, \int_{k_1}\hspace{-6mm}\mbox{$\sum$} \ \ \int_{k_2}\hspace{-6mm}\mbox{$\sum$} \ \ \int_{k_3}\hspace{-6mm}\mbox{$\sum$} \ \  V_\RR(k_1,k_2,k_3) \\
& \hspace{4cm} \times 
\langle \ttc^{k_3}, \sgreen(\ttc^{p_1})\rangle_\RR \, \langle \ttc^{k_2}, \sgreen(\ttc^{p_2})\rangle_\RR \, 
\langle \ttc^{k_1}, \sgreen(\ttc^{p_3})\rangle_\RR \\[4pt]
&=  (\ii\,\hbar)^2\, \Big(-\frac{g}{3!}\Big) \, (2\pi)^3 \  \delta(E_{1}+E_{2}+E_{3}) \ \delta(p_{1z}+p_{2z} +p_{3z} ) \ \delta_{\ell_{1}+\ell_{2}+\ell_{3}, 0} \\
& \hspace{3cm} \times  \widetilde{\sgreen}(p_1) \, \widetilde{\sgreen}(p_2) \, \widetilde{\sgreen}(p_3) \ (-1)^{\ell_{1}+\ell_{2}+\ell_{3}} \ \e^{\frac{\ii\, \lambda}{2}\, \sum\limits_{a<b}\, (p_{az}\, \ell_b - p_{bz}\, \ell_a)} \\
& \hspace{6cm} \times \ttF_{\ell_{1},\ell_{2},\ell_{3}} (\alpha_{1}, \alpha_{2}, \alpha_{3}) \ .\label{G3TreeTypical}
\end{split}
\end{align}

According to \eqref{3Bessel} and \eqref{2BesselAdd}, the integral of three cylindrical Bessel functions reduces to
\begin{align} \label{eq:Fellalphaexplicit}
\begin{split}
\ttF_{\ell_{1},\ell_{2},\ell_{3}} (\alpha_{1}, \alpha_{2}, \alpha_{3})  = \begin{cases}
 \ 0 \ , & \quad \vec\alpha_1,\vec\alpha_2,\vec\alpha_3\neq\vec0 \ , \ \vec\alpha_1+\vec \alpha_2+\vec \alpha_3\neq\vec 0 \ , \\[4pt]
 \ \displaystyle (-1)^{\ell_{3}}\,\frac{\cos \big(\ell_{1}\,\upsilon_{2} - \ell_{2}\,\upsilon_{1})}{\pi \,\alpha_{1}\,\alpha_{2} \sin (\upsilon_{3})} \ , & \quad  \vec\alpha_1,\vec\alpha_2,\vec\alpha_3\neq\vec0 \ , \ \vec\alpha_1+\vec \alpha_2+\vec \alpha_3=\vec 0 \ , \\[12pt]
 \ \displaystyle \delta_{\ell_1,0} \ (-1)^{\ell_3} \, \frac{\delta(\alpha_2-\alpha_3)}{\alpha_3} \ , & \quad \vec\alpha_1=\vec0  \ . 
\end{cases}
\end{split}
\end{align}
Thus there are special regions in momentum space where the interactions are non-trivial only when the conservation of momentum is fully restored: $p_1+p_2+p_3=0$. These are the triangular regions where all two-dimensional vectors $\vec\alpha_i$ are non-zero and the  radial momentum is conserved, as in \eqref{3momenta}; the triangle identities of Appendix~\ref{app:Bessel} ensure the symmetries \eqref{eq:Fellalphasym} even though they are not manifest in \eqref{eq:Fellalphaexplicit}. Otherwise, when $\vec\alpha_i=\vec0$ for some $i\in\{1,2,3\}$, only the magnitude of the radial momentum is conserved.

The final result for (\ref{G3TreeTypical}) is given by
\begin{align}
\begin{split}
\widetilde C_3^\RR(p_1, p_2,p_3)^{\swzero}_{\1PI} &= (\ii\,\hbar)^2\, \Big(-\frac{g}{3!}\Big) \, (2\pi)^3 \ \delta(E_{1}+E_{2}+E_{3}) \ \delta(p_{1z}+p_{2z} +p_{3z} ) \ \delta_{\ell_{1} + \ell_{2} + \ell_{3}, 0}  \\
& \hspace{1cm} \times  \widetilde{\sgreen}(p_1) \, \widetilde{\sgreen}(p_2) \, \widetilde{\sgreen}(p_3) \ \e^{-\frac{\ii \, \lambda}{2}\, (p_{1z}\, \ell_{2} - p_{2z}\, \ell_{1} )} \ \ttF_{\ell_{1},\ell_{2},\ell_{3}} (\alpha_{1}, \alpha_{2}, \alpha_{3}) \ . \label{G3TreeTypicalResult2}
\end{split}
\end{align}
The noncommutative contribution simply appears in the form of a phase factor.

\subsection{Propagator corrections at one-loop}%
\label{sub:1loop2ptbraided}

Before turning to  explicit calculations at loop level, we outline the strategy followed through this
subsection. First we sum over loop angular momenta which will force conservation of incoming radial momenta. Because the angular twist breaks translational invariance in the
$(x,y)$-plane, the radial momentum $\alpha$ is not on the same footing as the energy $E$
and the axial momentum $k_z$, so the usual trick of Wick rotating and passing to
hyperspherical coordinates in all four momentum components is not handy. Instead we
combine $E$ and $k_z$ into a two-dimensional vector $R = (E, k_z)$, apply a Feynman
parametrisation to the resulting integral, and treat the radial momentum integration
separately using the Bessel function identities of Appendix \ref{app:Bessel}. To extract the
ultraviolet behaviour in closed form, we specialise to external states with purely
axial momentum, $\alpha_1 = 0$, and show in (\ref{UVStep4}) that the familiar
logarithmic ultraviolet divergence of $\varPhi^3$-theory in four dimensions appears. We then check this
explicitly against the standard one-loop calculation performed directly in the plane
wave basis. The same strategy is applied to the standard (unbraided) noncommutative
theory in Section \ref{sub:2point1loop}.

The one-loop contribution to the two-point function $\widetilde C_2^\RR(p_1, p_2)$ is given by
\begin{align}
\begin{split}
\widetilde C_2^\RR(p_1, p_2)^{\swone} &=  \ii\,\hbar\,\BVL\,\sH \,\bigl\{ \CS^\RR _{\rm int},\sH\, \big(\ii\,\hbar\,\BVL\,\sH\,\bigl\{\CS^\RR _{\rm int},\sH\,(\ttc^{p_1}\odot_\RR \ttc^{p_2})\bigr\}_\RR \big)\bigr\}_\RR \\
& \quad \, + (\ii\,\hbar\,\BVL\,\sH)^2 \, \bigl\{ \CS^\RR _{\rm int},\sH\, \bigl\{\CS^\RR _{\rm int},\sH\,(\ttc^{p_1}\odot_\RR \ttc^{p_2})\bigr\}_\RR \bigr\}_\RR \ .\label{TwoPoint}
\end{split}
\end{align}
The explicit calculations repeat those of \cite{Bogdanovic:2024jnf} exactly, where a diagrammatic calculus was developed. Applying this,  the complete result for the one-loop two-point function is given by
\begin{eqnarray}\label{KorekcijaPhi3Diagrams}
\widetilde C_2^\RR(p_1,p_2)^{\swone} = { 9 \times { \footnotesize \begin{tikzpicture}[scale=0.5, baseline]
        \coordinate (k) at (-1.5,0);
        \coordinate (l) at (1.0,0);
        \coordinate[label=above: $p_1$] (p1) at (180:2.5);
        \coordinate[label=above: $p_2$] (p2) at (0:2.0);
        \draw[decoration={markings, mark=at position 0.5 with {\arrow{Latex[reversed]}} }, postaction={decorate}] (k) -- (p1);
        \draw[decoration={markings, mark=at position 0.5 with {\arrow{latex[reversed]}} }, postaction={decorate}] (l) -- (p2);
        \draw[decoration={markings}, postaction={decorate}] ($(k)+(0:0.5)$) circle (0.5);
        \draw[decoration={markings}, postaction={decorate}] ($(l)-(0:0.5)$) circle (0.5);
    \end{tikzpicture} } \normalsize
+18 \times { \footnotesize \begin{tikzpicture}[scale=0.5, baseline]
        \coordinate (k) at (0,0);
        \coordinate (l) at ($(k)+(90:1)$);
        \coordinate[label=above: $p_1$] (p1) at ($(k) +(180:2.5)$);
        \coordinate[label=above: $p_2$] (p2) at ($(k) +(0:2.5)$);

        \draw[decoration={markings, mark=at position 0.5 with {\arrow{Latex[reversed]}} }, postaction={decorate}] (k) -- (p1);
        \draw[decoration={markings, mark=at position 0.5 with {\arrow{Latex[reversed]}} }, postaction={decorate}] (k) -- (p2);
        \draw[decoration={markings}, postaction={decorate}] (k) -- (l);
        \draw[decoration={markings}, postaction={decorate}] ($(l)+(90:0.5)$) circle (0.5);
    \end{tikzpicture} } \normalsize
+ 18 \times { \footnotesize \begin{tikzpicture}[scale=0.5, baseline]
        \coordinate (k) at (-0.5,0);
        \coordinate (l) at (0.5,0);
        \draw[decoration={markings, mark=at position 0.5 with {\arrow{Latex[reversed]}} }, postaction={decorate}] (k) -- ($(k) + (180:2)$) node[above]{$p_1$};
        \draw[decoration={markings, mark=at position 0.5 with {\arrow{Latex[reversed]}} }, postaction={decorate}] (l) -- ($(l) + (0:2)$) node[above]{$p_2$};
        \draw[decoration={markings }, postaction={decorate}] ($(k)+(0:0.5)$) circle (0.5);    
    \end{tikzpicture} } \normalsize
   }  \ .
\end{eqnarray}

Each of the diagrammatic contributions in \eqref{KorekcijaPhi3Diagrams} can be treated individually, using the explicit formulas from~\cite{Bogdanovic:2024jnf}. We will discard the tadpole contributions; they can be eliminated by curving the $L_\infty$-structure as explained in~\cite{Bogdanovic:2024jnf}.
 In the following we label the components of internal loop momenta explicitly by their Latin symbol for clarity, whereas external momentum components will remain without this designation.

\paragraph{\underline{\sf One-particle irreducible contribution.}}

Consider the contribution to (\ref{KorekcijaPhi3Diagrams}) from the one-particle irreducible diagram
\begin{align} \label{IIiIII}
\begin{split}
& \hspace{-3mm} { \footnotesize { \begin{tikzpicture}[scale=0.5, baseline]
        \coordinate (k) at (-0.5,0);
        \coordinate (l) at (0.5,0);
        \draw[decoration={markings, mark=at position 0.5 with {\arrow{Latex[reversed]}} }, postaction={decorate}] (k) -- ($(k) + (180:2)$) node[above]{$p_1$};
        \draw[decoration={markings, mark=at position 0.5 with {\arrow{Latex[reversed]}} }, postaction={decorate}] (l) -- ($(l) + (0:2)$) node[above]{$p_2$};
        \draw[decoration={markings}, postaction={decorate}] ($(k)+(0:0.5)$) circle (0.5);    
    \end{tikzpicture} } \normalsize }\\[4pt]
&  \hspace{-3mm} \quad = \frac{1}{3} \, \frac{1}{2} \, (\ii\,\hbar)^2\, \Big(-\frac{g}{3!}\Big)^2\, (-1)^{\ell_{2}}\, (2\pi)^3 \ \delta(E_{1}+E_{2}) \ \delta(p_{1z}+p_{2z}) \ \delta_{\ell_{1}+\ell_{2}, 0} \ \widetilde{\sgreen}(p_1) \ \widetilde{\sgreen}(p_2)\\
& \hspace{-3mm}\hspace{1cm} \times  \ \int_{q_1}\hspace{-5.8mm}\mbox{$\sum$} \ \  \widetilde{\sgreen}(q_1) \ \int_0^\infty\, \alpha_{q_2} \, \d \alpha_{q_2} \ \frac{\ttF_{\ell_{1} + \ell_{q_1},-\ell_{q_1},-\ell_{1}} (\alpha_{q_2}, \alpha_{q_1}, \alpha_{1}) \, \ttF_{\ell_{q_1},-\ell_{2},-\ell_{1}-\ell_{q_1}} (\alpha_{q_1}, \alpha_{2}, \alpha_{q_2}) }{\alpha^2_{q_2} - (E_{1} + E_{q_1})^2 + (p_{1z} + q_{1z})^2 + m_+^2} \ .    
\end{split}
\end{align}
The evaluation of the loop integrals in \eqref{IIiIII} is subtle due to the discontinuous behaviour of \eqref{eq:Fellalphaexplicit}. We  assume that $\alpha_i\neq0$ and restrict the radial loop momentum integrals to the open intervals $\alpha_{q_i}\in(0,\infty)$.
The integrals of Bessel functions then give the contributions
\begin{align}\label{eq:1PIBesselint}
\begin{split}
\ttF_{\ell_{1} + \ell_{q_1},-\ell_{q_1},-\ell_{1}} (\alpha_{q_2}, \alpha_{q_1}, \alpha_{1}) &= 
(-1)^{\ell_{1}+\ell_{q_1}}\,\frac{\cos(\ell_{1}\,\upsilon_{q_1}-\ell_{q_1}\,\upsilon_{1})}{\pi\,\alpha_{1}\,\alpha_{q_1} \sin(\upsilon_{q_2})}\ ,
\\[4pt]
\ttF_{\ell_{q_1},\ell_{1},-\ell_{1}-\ell_{q_1}} (\alpha_{q_1}, \alpha_{2}, \alpha_{q_2}) &= 
(-1)^{\ell_{1}+\ell_{q_1}} \, \frac{\cos(\ell_{1}\,\upsilon'_{q_1}-\ell_{q_1}\,\upsilon_{2})}{\pi\,\alpha_{2}\,\alpha_{q_1} \sin(\upsilon'_{q_2})}\ .
\end{split}
\end{align}

We start by evaluating the sum over $\ell_{q_1}$ using Euler's formula to obtain
\begin{align}\label{Ireducible:SumOverLq1}
    \frac{1}{2\pi}\,\sum_{\ell_{q_1}\in\RZ} \, \cos(\ell_1\,\upsilon_{q_1}-\ell_{q_1}\,\upsilon_1) \cos(\ell_1\,\upsilon'_{q_1}-\ell_{q_1}\,\upsilon_2) &= \frac{1}{2}\,\Big(\delta(\upsilon_1 - \upsilon_2) \cos\big(\ell_1\,(\upsilon_{q_1}-\upsilon_{q_1}')\big) \nn \\[-12pt]
    & \hspace{2cm} + \delta(\upsilon_1 + \upsilon_2)\cos\big(\ell_1\,(\upsilon_{q_1}+\upsilon_{q_1}')\big)\Big) \ .\nn
\end{align}
Since by definition $\upsilon_i$ are interior angles of a triangle and cannot be negative, the second Dirac distribution can be discarded. For the first Dirac distribution, we exploit triangle relations \eqref{3Bessel} to swap angles with sides. The angles $\upsilon_1$ and $\upsilon_2$ are monotonic functions of the radial momenta $\alpha_1$ and $\alpha_2$, respectively. This implies that $\upsilon_1-\upsilon_2$ can be treated as a function of $\alpha_1$, and $\upsilon_1=\upsilon_2$ if and only if $\alpha_1=\alpha_2$. 

The cosine law applied to the first momentum triangle in \eqref{eq:1PIBesselint} gives
\begin{equation}
    \cos(\upsilon_1) = \frac{\alpha^2_{q_1} + \alpha^2_{q_2} - \alpha^2_1}{2\,\alpha_{q_1}\, \alpha_{q_2}} \ .\nn
\end{equation}
Taking the derivative with respect to $\alpha_1$ gives
\begin{equation}
     \sin(\upsilon_1)\,\frac{\dd\upsilon_1}{\dd\alpha_1} = \frac{\alpha_1}{\alpha_{q_1}\,\alpha_{q_2}} \ .\nn
\end{equation}
The expected radial momentum conservation law now follows as
\begin{equation}\label{Ireducible:DeltaToDelta}
    \delta(\upsilon_1 - \upsilon_2) = \frac{\delta(\alpha_1-\alpha_2)}{\frac{\dd\upsilon_1}{\dd\alpha_1}} = \alpha_{q_1}\,\alpha_{q_{2}} \sin(\upsilon_1)\,\frac{\delta(\alpha_1 - \alpha_2)}{\alpha_1} \ .
\end{equation}
There is an additional consequence of \eqref{Ireducible:DeltaToDelta}: both momentum triangles in \eqref{eq:1PIBesselint} now have equal sides, and therefore they have equal angles, i.e. $\upsilon'_{q_i} = \upsilon_{q_i}$ for $i=1,2$.

Applying sine and cosine laws we may now rewrite (\ref{IIiIII}) as
\begin{align}
\begin{split}
&{ \footnotesize { \begin{tikzpicture}[scale=0.5, baseline]
        \coordinate (k) at (-0.5,0);
        \coordinate (l) at (0.5,0);
        \draw[decoration={markings, mark=at position 0.5 with {\arrow{Latex[reversed]}} }, postaction={decorate}] (k) -- ($(k) + (180:2)$) node[above]{$p_1$};
        \draw[decoration={markings, mark=at position 0.5 with {\arrow{Latex[reversed]}} }, postaction={decorate}] (l) -- ($(l) + (0:2)$) node[above]{$p_2$};
        \draw[decoration={markings}, postaction={decorate}] ($(k)+(0:0.5)$) circle (0.5);    
\end{tikzpicture} } \normalsize } \\[4pt]
& \quad = \frac{4\pi}{3} \, (\ii\,\hbar)^2\, \Big(-\frac{g}{3!}\Big)^2\, (-1)^{\ell_{2}} \ \delta(E_{1}+E_{2}) \ \frac{\delta(\alpha_1 - \alpha_2)}{\alpha_1} \ \delta(p_{1z}+p_{2z}) \ \delta_{\ell_{1}+\ell_{2}, 0} \\
&\hspace{1cm} \times \widetilde{\sgreen}(p_1) \, \widetilde{\sgreen}(p_2) \,\int_{\FR\times\FR}\, \frac{\d E_{q_1}}{2\pi} \ \frac{\d q_{1z}}{2\pi} \ \int_{0^+}^\infty\, \alpha_{q_1}\,\d \alpha_{q_1} \ \widetilde{\sgreen}(q_1) \label{Irreducible} \\
& \hspace{1.3cm} \times \int_{0^+}^\infty\,   \frac{\alpha_{q_2}\,\d \alpha_{q_2}}{\alpha^2_{q_2} - (E_{1} + E_{q_1})^2 + (p_{1z} + q_{1z})^2 + m_+^2 } \ \frac{2}{\sqrt{4\,\alpha^2_{q_1}\,\alpha^2_{q_2}-(\alpha^2_{q_1} + \alpha^2_1 - \alpha^2_{q_2} )^2}} \ .
\end{split}
\end{align}

\paragraph{\underline{\sf Ultraviolet behaviour.}}

Let us now discuss the ultraviolet behaviour of the one-particle irreducible two-point function at one-loop order (\ref{Irreducible}). It is well-known that the corresponding commutative result is logarithmically ultraviolet divergent when computed in the standard plane wave basis. As there are no noncommutative contributions in (\ref{Irreducible}), it should also exhibit some remnant of this ultraviolet divergence, as such infinities cannot be removable by simply passing to another basis for the spectral decomposition of the scalar fields. In the following we show how the standard ultraviolet divergences are exhibited in the cylindrical harmonic basis.

The internal linear momentum along the $z$-direction and radial momentum are not on equal footing, as a consequence of the breaking of translational symmetry in the $(x,y)$-plane. Hence we cannot apply the standard method of Wick rotating the time direction and changing variables to hyperspherical coordinates in momentum space, in which the radial coordinate corresponds to the magnitude of the total momentum. Instead, we analyse the ultraviolet behaviour by combining the axial momentum with the energy, and then incorporating the radial momentum. 

We set $ R=(E,p_z)$ and start from the integral
\begin{align}
\begin{split}
& \int_{\FR\times\FR} \,\frac{{\rm d}E_{q_1} \, {\rm d}q_{1z}}{\big(\alpha_{q_1}^2-E_{q_1}^2  +q_{1z}^2 + m_+^2 \big)  \big(\alpha^2_{q_2} - (E_{1} + E_{q_1})^2 + (p_{1z} + q_{1z})^2 + m_+^2 \big)} \\[4pt]
& \hspace{1cm}= \int_{\FR^2}\,  \frac{{\rm d}^2 R_{q_1} }{\big(R_{q_1}^2 - \alpha_{q_1}^2 - m_+^2 \big) \, \big(({R}_{1} + {R}_{q_1})^2 -\alpha^2_{q_2} - m_+^2 \big)} \\[4pt]
& \hspace{1cm} = \int_0^1 {\rm d}x \ \int_{\FR^2}\, \frac{{\rm d}^2 R_{q_1} }{\big( R_{q_1}^2 + 2\,(1-x)\,{R}_{q_1}\cdot {R}_{1} - m_+^2 - x\, \alpha^2_{q_1} - (1-x)\,(\alpha^2_{q_2} - R_1^2) \big)^2} \\[4pt]
& \hspace{1cm} = \ii\, \pi\, \int_0^1\, {\rm d}x \ \frac{1}{x\,\alpha^2_{q_1} + (1-x)\,\alpha^2_{q_2} + m_+^2 -x\,(1-x)\,R_1^2} \ . \label{UVStep1}
\end{split}
\end{align}
In the second equality we introduced the standard Feynman parametrisation, while
the last equality is obtained using the table of $D$-dimensional integrals in Minkowski space from \cite[eq.~(11A)]{Radovanovic:2008zz} for~$D=2$. 

The final expression for the irreducible contribution is then
\begin{align}
\begin{split}
&{ \footnotesize { \begin{tikzpicture}[scale=0.5, baseline]
        \coordinate (k) at (-0.5,0);
        \coordinate (l) at (0.5,0);
        \draw[decoration={markings, mark=at position 0.5 with {\arrow{Latex[reversed]}} }, postaction={decorate}] (k) -- ($(k) + (180:2)$) node[above]{$p_1$};
        \draw[decoration={markings, mark=at position 0.5 with {\arrow{Latex[reversed]}} }, postaction={decorate}] (l) -- ($(l) + (0:2)$) node[above]{$p_2$};
        \draw[decoration={markings}, postaction={decorate}] ($(k)+(0:0.5)$) circle (0.5);    
\end{tikzpicture} } \normalsize } \\[4pt]
& \hspace{1cm} = \frac{\ii}{3} \, (\ii\,\hbar)^2\, \Big(-\frac{g}{3!}\Big)^2\, (-1)^{\ell_{2}} \ \delta(E_{1}+E_{2})  \ \frac{\delta(\alpha_1 - \alpha_2)}{\alpha_1} \ \delta(p_{1z}+p_{2z}) \ \delta_{\ell_{1}+\ell_{2}, 0} \\
&\hspace{2cm} \times \widetilde{\sgreen}(p_1) \, \widetilde{\sgreen}(p_2) \,\int_{0^+}^\infty\,\alpha_{q_1}\,\d \alpha_{q_1} \ \int_{0^+}^\infty\, \alpha_{q_2}\,\d \alpha_{q_2} \ \frac{2}{\sqrt{4\,\alpha^2_{q_1}\alpha^2_{q_2}-(\alpha^2_{q_1} + \alpha^2_1 - \alpha^2_{q_2} )^2}} \label{Irreducible1}\\
&\hspace{3cm}\times \int_0^1\,\dd x \ \frac{1}{x\,\alpha^2_{q_1} + (1-x)\,\alpha^2_{q_2} + m_+^2 -x\,(1-x)\,(E_1^2-p_{1z}^2)}  \ .
\end{split}
\end{align}
The integrations in \eqref{Irreducible1} are symmetric under exchange  $\alpha_{q_1}\leftrightarrow\alpha_{q_2}$ together with $x \rightarrow 1-x$. The integral over $\alpha_{q_1}$ can now be performed. The full analysis is somewhat technically involved and not very informative, so we will consider a simplifying instance.

In the special case where one of the external momenta is purely axial, say $\alpha_1 = 0$, the radial integrations in \eqref{Irreducible1} can be made explicit, prior to introducing the Feynman parametrisation. When $\alpha_1=0$ the first integral of Bessel functions in \eqref{eq:1PIBesselint} reduces according to \eqref{eq:Fellalphaexplicit} as
\begin{align*}
    \ttF_{\ell_{1} + \ell_{q_1},-\ell_{q_1},-\ell_{1}} (\alpha_{q_2}, \alpha_{q_1},0) = \delta_{\ell_1,0} \ (-1)^{\ell_{q_1}} \, \frac{\delta(\alpha_{q_1} - \alpha_{q_2})}{\alpha_{q_1}}  \ .
\end{align*}
The second integral of Bessel functions in \eqref{eq:1PIBesselint} can be summed over $\ell_{q_1}\in\RZ$, which leads to the conservation of radial momenta in the same manner as in (\ref{Ireducible:DeltaToDelta}). 
Altogether this results in
\begin{align*}
\begin{split}
\hspace{-3mm} { \footnotesize { \begin{tikzpicture}[scale=0.5, baseline]
        \coordinate (k) at (-0.5,0);
        \coordinate (l) at (0.5,0);
        \draw[decoration={markings, mark=at position 0.5 with {\arrow{Latex[reversed]}} }, postaction={decorate}] (k) -- ($(k) + (180:2)$) node[above]{$p_1$};
        \draw[decoration={markings, mark=at position 0.5 with {\arrow{Latex[reversed]}} }, postaction={decorate}] (l) -- ($(l) + (0:2)$) node[above]{$p_2$};
        \draw[decoration={markings}, postaction={decorate}] ($(k)+(0:0.5)$) circle (0.5);    
    \end{tikzpicture} } \normalsize } \ \Big|_{\alpha_1=0}  &= \frac{\ii\, \pi^2}{3} \, (\ii\,\hbar)^2\, \Big(-\frac{g}{3!}\Big)^2 \ \delta(E_{1}+E_{2}) \ \delta(p_{1z}+p_{2z}) \ \delta_{\ell_{1}, 0} \ \delta_{\ell_{2}, 0} \ \widetilde{\sgreen}(p_1) \ \widetilde{\sgreen}(p_2) \\
& \qquad \, \times \int_{0^+}^\infty\, \alpha_{q_1} \, \d \alpha_{q_1} \ \int_0^1 {\rm d}x \ \frac{ \alpha_{q_1}\,\delta(\alpha_2)}{\pi\, \alpha_{q_1}\,\alpha_2\,\big(\alpha^2_{q_1} + m_+^2 - x\,(1-x)\,R_1^2\big)} \ . 
\end{split}
\end{align*}

The integration over $\alpha_{q_1}\in(0,\Lambda]$ may now be performed explicitly, where $\Lambda\to\infty$ is an ultraviolet cutoff, and subsequently over $x\in[0,1]$. The final result is
\begin{align}
    \begin{split}
& \hspace{-3mm} { \footnotesize { \begin{tikzpicture}[scale=0.5, baseline]
        \coordinate (k) at (-0.5,0);
        \coordinate (l) at (0.5,0);
        \draw[decoration={markings, mark=at position 0.5 with {\arrow{Latex[reversed]}} }, postaction={decorate}] (k) -- ($(k) + (180:2)$) node[above]{$p_1$};
        \draw[decoration={markings, mark=at position 0.5 with {\arrow{Latex[reversed]}} }, postaction={decorate}] (l) -- ($(l) + (0:2)$) node[above]{$p_2$};
        \draw[decoration={markings}, postaction={decorate}] ($(k)+(0:0.5)$) circle (0.5);    
    \end{tikzpicture} } \normalsize } \ \Big|_{\alpha_1=0} \\[4pt]
    & \hspace{1cm} = \frac{\ii\, \pi}{6} \, (\ii\,\hbar)^2\, \Big(-\frac{g}{3!}\Big)^2 \ \delta(E_{1}+E_{2}) \ \frac{\delta(\alpha_2)}{\alpha_2} \ \delta(p_{1z}+p_{2z}) \ \delta_{\ell_{1}, 0} \ \delta_{\ell_{2}, 0} \ \widetilde{\sgreen}(p_1) \ \widetilde{\sgreen}(p_2) \\
& \hspace{1.5cm} \times    \Bigg(\log\bigg(\frac{\Lambda^2}{m^2}+1\bigg) +\sqrt{\frac{4\,(\Lambda^2+m^2)}{E_1^2-p_{1z}^2}-1} \ \tan^{-1}\sqrt{\frac{E_1^2-p_{1z}^2}{4\,(\Lambda^2+m^2)-(E_1^2-p_{1z}^2)}} \\
& \hspace{6.5cm} - \sqrt{\frac{4m^2}{E_1^2-p_{1z}^2}-1} \ \tan^{-1}\sqrt{\frac{E_1^2-p_{1z}^2}{4m^2-(E_1^2-p_{1z}^2)}} \ \Bigg) \ ,    \label{UVStep4}
\end{split}
\end{align}
where we use the representation  of the arctangent function 
$$\tan^{-1}(w)=\frac1{2\,\ii}\log\bigg(\frac{1+\ii\,w}{1-\ii\,w}\bigg)$$ which is valid for $w\in\FC\setminus\{\pm\,\ii\}.$ 

The expression \eqref{UVStep4} exhibits two important properties. Firstly, we observe that the Dirac distribution in the external radial momentum  exactly fixes $\alpha_2=0$, as expected by momentum conservation and since $\alpha_1=0$ here; as also anticipated both external angular momenta vanish: $\ell_1=\ell_2=0$. Secondly, in (\ref{UVStep4}) we immediately recognize the logarithmic ultraviolet divergence characteristic of the commutative $\varPhi^3$-theory in four dimensions. 

\paragraph{\underline{\sf Standard one-loop correction.}}

To compare \eqref{UVStep4}  with its commutative counterpart, and in particular to show that it reproduces the same logarithmic ultraviolet divergence, we recalculate the standard one-loop correction to the propagator obtained in the plane wave basis via the usual Feynman rules. It is given by
\begin{align}
\begin{split}\label{StandarLogDiv}
& \hspace{-3mm} { \footnotesize { \begin{tikzpicture}[scale=0.5, baseline]
        \coordinate (k) at (-0.5,0);
        \coordinate (l) at (0.5,0);
        \draw[decoration={markings, mark=at position 0.5 with {\arrow{Latex[reversed]}} }, postaction={decorate}] (k) -- ($(k) + (180:2)$) node[above]{$p_1$};
        \draw[decoration={markings, mark=at position 0.5 with {\arrow{Latex[reversed]}} }, postaction={decorate}] (l) -- ($(l) + (0:2)$) node[above]{$p_2$};
        \draw[decoration={markings}, postaction={decorate}] ($(k)+(0:0.5)$) circle (0.5);    
    \end{tikzpicture} } \normalsize } = \frac1{3} \, (\ii\,\hbar)^2\, \Big(-\frac{g}{3!}\Big)^2\, (2\pi)^4 \ \delta(p_{1}+p_{2}) \ \widetilde{\sgreen}(p_1) \ \widetilde{\sgreen}(p_2)\\
& \hspace{-3mm}\hspace{8cm} \times \int_{\FR^4}\, \frac{\d^4 q}{(2\pi)^4} \ \frac{1}{q^2-m_+^2}\,\frac{1}{(p_1-q)^2-m_+^2}\ .
\end{split}
\end{align}
After Feynman parametrisation, (\ref{StandarLogDiv}) reduces to
\begin{align}
\begin{split} \label{StandarLogDiv2}
&{ \footnotesize { \begin{tikzpicture}[scale=0.5, baseline]
        \coordinate (k) at (-0.5,0);
        \coordinate (l) at (0.5,0);
        \draw[decoration={markings, mark=at position 0.5 with {\arrow{Latex[reversed]}} }, postaction={decorate}] (k) -- ($(k) + (180:2)$) node[above]{$p_1$};
        \draw[decoration={markings, mark=at position 0.5 with {\arrow{Latex[reversed]}} }, postaction={decorate}] (l) -- ($(l) + (0:2)$) node[above]{$p_2$};
        \draw[decoration={markings}, postaction={decorate}] ($(k)+(0:0.5)$) circle (0.5);    
    \end{tikzpicture} } \normalsize } = \frac13\,(\ii\,\hbar)^2\, \Big(-\frac{g}{3!}\Big)^2\, (2\pi)^4 \ \delta(p_{1}+p_{2}) \ \widetilde{\sgreen}(p_1) \ \widetilde{\sgreen}(p_2)\\
& \hspace{5cm} \times \int_{\FR^4}\, \frac{\d^4 q}{(2\pi)^4} \ \int_0^1\, {\rm d}x \ \frac{1}{\big(x\,q^2+(1-x)\,(p_1-q)^2-m_+^2\big)^2}\ .   
\end{split}
\end{align}

Changing to cylindrical coordinates in momentum space results in the four-dimensional Dirac distribution
\begin{equation}
\delta(p_{1}+p_{2}) = \delta(E_{1}+E_{2}) \ \frac{\delta(\alpha_{1}-\alpha_{2})}{\alpha_1}  \ \delta(\vartheta_1-\vartheta_2-\pi) \ \delta(p_{1z}+p_{2z}) \ , \nn
\end{equation}
where $\vartheta_i$ are polar angles in the $(x,y)$-plane of momentum space (see Appendix~\ref{app:Bessel}).
The denominator in (\ref{StandarLogDiv2}) can be cast into the form
\begin{equation}
\big(x\,q^2+(1-x)\,(p_1-q)^2-m_+^2\big)^2 = \big({R}_q^2 - 2\,(1-x)\,{R}_1 \cdot {R}_q - \alpha_q^2 - m_+^2+(1-x)\,(R_1^2-\alpha_1^2)\big)^2 \nn
\end{equation}
where ${R_q} = (E_q, q_z)$ and ${R_1} = (E_1, p_{1z})$. The integral over ${R}_q\in\FR^2$ can be easily performed, as shown in (\ref{UVStep1}). Then
\begin{align}\label{StandarLogDiv3}
\begin{split}
{ \footnotesize { \begin{tikzpicture}[scale=0.5, baseline]
        \coordinate (k) at (-0.5,0);
        \coordinate (l) at (0.5,0);
        \draw[decoration={markings, mark=at position 0.5 with {\arrow{Latex[reversed]}} }, postaction={decorate}] (k) -- ($(k) + (180:2)$) node[above]{$p_1$};
        \draw[decoration={markings, mark=at position 0.5 with {\arrow{Latex[reversed]}} }, postaction={decorate}] (l) -- ($(l) + (0:2)$) node[above]{$p_2$};
        \draw[decoration={markings}, postaction={decorate}] ($(k)+(0:0.5)$) circle (0.5);    
    \end{tikzpicture} } \normalsize } & = \frac13\,(\ii\,\hbar)^2\, \Big(-\frac{g}{3!}\Big)^2 \  \delta(E_{1}+E_{2}) \ \frac{\delta(\alpha_{1}-\alpha_{2})}{\alpha_1} \ \delta(\vartheta_1-\vartheta_2-\pi) \ \delta(p_{1z}+p_{2z})  \\
& \quad \, \times \widetilde{\sgreen}(p_1) \ \widetilde{\sgreen}(p_2) \ \int_0^1\, {\rm d}x \ \int_0^\infty\, \alpha_q \, \d \alpha_q  \ \frac{\ii\, \pi}{\alpha_q^2+m_+^2+(1-x)\,\big(\alpha_1^2-x\,{R}_1^2\big)} \ . 
\end{split}
\end{align}

Integrating over $\alpha_q\in[0,\Lambda]$ now reduces (\ref{StandarLogDiv3}) to
\begin{align}
\begin{split}
& { \footnotesize { \begin{tikzpicture}[scale=0.5, baseline]
        \coordinate (k) at (-0.5,0);
        \coordinate (l) at (0.5,0);
        \draw[decoration={markings, mark=at position 0.5 with {\arrow{Latex[reversed]}} }, postaction={decorate}] (k) -- ($(k) + (180:2)$) node[above]{$p_1$};
        \draw[decoration={markings, mark=at position 0.5 with {\arrow{Latex[reversed]}} }, postaction={decorate}] (l) -- ($(l) + (0:2)$) node[above]{$p_2$};
        \draw[decoration={markings}, postaction={decorate}] ($(k)+(0:0.5)$) circle (0.5);    
    \end{tikzpicture} } \normalsize } \\[4pt]
    & \hspace{1cm} = \frac{\ii\,\pi}{6} \, (\ii\,\hbar)^2\, \Big(-\frac{g}{3!}\Big)^2 \ \delta(E_{1}+E_{2}) \ \frac{\delta(\alpha_1 - \alpha_{2})}{\alpha_1} \ \delta(\vartheta_1-\vartheta_2-\pi) \ \delta(p_{1z}+p_{2z})  \\
& \hspace{2cm} \times \widetilde{\sgreen}(p_1) \ \widetilde{\sgreen}(p_2) \ \int_0^1\, {\rm d}x \ \log\bigg(\frac{\Lambda^2}{m^2+(1-x)\,\alpha_1^2-x\,(1-x)\,(E_1^2-p_{1z}^2)} + 1 \bigg) \ . \label{UVStep5}
\end{split}
\end{align}
After integration over $x\in[0,1]$, the correction (\ref{UVStep5}) at $\alpha_1=0$ is the same as (\ref{UVStep4}), confirming that the standard logarithmic divergence is reproduced by our braided one-loop two-point function evaluated in the basis of Bessel functions.

\paragraph{\underline{\sf Absence of non-planar contributions.}}

The calculations of this section, going back to the basic form (\ref{IIiIII}), show that there is no noncommutative contribution to the two-point function (\ref{TwoPoint}). The braided vertex (\ref{vertex2}) and the Braided Wick Theorem combine in such a way as to cancel all non-planar contributions, leaving the same result as the corresponding calculations in the undeformed theory (performed in the basis of cylindrical harmonics), up to phase factors which depend only on the momenta of external states. This seems to be a general feature of braided BV-quantized field theories in the absence of non-abelian gauge symmetries~\cite{Nguyen:2021rsa,DimitrijevicCiric:2023hua,Bogdanovic:2024jnf,DimitrijevicCiric:2024qew}.

\section{Standard Batalin-Vilkovisky formalism for  $\mbf{\varPhi_\star^3}$-theory} \label{StarQFT}%

The braided $\varPhi_\star^3$ quantum field theory of Section~\ref{BraidedQFT} was constructed by applying the twist deformation to the  $L_\infty$-algebra of the corresponding commutative theory, thus intrinsically introducing noncommutativity. However, this is not the only possible noncommutative deformation of the homotopy algebra organising the commutative model~\cite{Giotopoulos:2021ieg}. In the following we present another version based on an ordinary (unbraided) $L_\infty$-algebra, which leads to the standard noncommutative $\varPhi^3_\star$ quantum field theory. Following the discussion at the end of Section~\ref{QFT}, one can regard \eqref{braidedell} as the brackets of an $A_\infty$-algebra (in fact a differential graded noncommutative algebra) in the category of vector spaces $\CCV$, whose graded antisymmetrization in $\CCV$ defines an ordinary $L_\infty$-structure. The computation of quantum correlation functions of the corresponding commutative theory directly in terms of this $A_\infty$-structure is discussed in~\cite{Okawa:2022sjf}.

Standard noncommutative scalar quantum field theories have been discussed in e.g.~\cite{Minwalla:1999px, Douglas:2001ba, Szabo:2001kg} for the Moyal deformation and in \cite{DimitrijevicCiric:2018blz, Hersent:2023lqm} for the case of $\lambda$-Minkowski space, using the conventional basis of plane waves and Feynman quantization. These theories involve both planar and non-planar contributions to one-loop correlation functions, and UV/IR mixing is present. In the following we present the details of the algebraic BV formalism that reproduces the same quantitative results.

\subsection{$L_\infty$-structure}

The $L_\infty$-algebra $(V, \mu_1,\mu_2)$ corresponding to the standard noncommutative theory again consists of the graded vector space $$V=V^1\oplus V^2$$ with physical fields $\varPhi\in V^1=\CCC^\infty(\FR^4)$ and their antifields $\varPhi^+\in V^2=\CCC^\infty(\FR^4)$.  The non-vanishing brackets are given by 
\begin{align} \label{unbraidedell}
\begin{split}
\mu_1 (\varPhi) &= \big(-\square -m^2\big)\, \varPhi \ , \\[4pt]
\mu_2 (\varPhi_1, \varPhi_2) &= \tfrac{1}{2} \, g\ (\varPhi_1\star\varPhi_2 + \varPhi_2\star\varPhi_1) = \mu_2 (\varPhi_2, \varPhi_1) \ .
\end{split}
\end{align}
Compared to the braided $L_\infty$-structure (\ref{braidedell}), the binary bracket $\mu_2$ is strictly symmetric, and so \eqref{unbraidedell} defines a strict $L_\infty$-structure in the category of vector spaces $\CCV$. Noncommutativity is now implicitly introduced through the symmetrization of the star-product in (\ref{unbraidedell}), rather than explicitly through a braiding as before. 

This $L_\infty$-algebra is strictly cyclic, with the same cyclic pairing (\ref{CyclPairing}):
\begin{equation}\label{CyclPairing_2}
\langle \varPhi, \varPhi ^+\rangle := \int_{\FR^4}\,\d^4 x \ \varPhi \cdot \varPhi^+  \ .
\end{equation}
Since $\mu_1 = \mu_1^\RR$ and $\mu_2(\varPhi,\varPhi) = \mu_2^\RR(\varPhi,\varPhi)$, the Maurer-Cartan functional coincides again with the classical action~\eqref{S_class}:
\begin{equation}
    S_{\rm cl} = \tfrac{1}{2!} \, \langle \varPhi, \mu_1(\varPhi)\rangle - \tfrac{1}{3!} \, \langle\varPhi, \mu_2(\varPhi, \varPhi)\rangle \ . \nn 
\end{equation}
In particular, the cohomology $H^\bullet(V)$ of the cochain complex $(V,\mu_1)$ is also the same as before.

Unlike the braided theory discussed in  Section~\ref{BraidedQFT}, in this setting there are no restrictions of equivariance on the projection map or the contracting homotopy in $\CCV$, and therefore no preferred basis of fields or coordinate frame to work in. Nevertheless, we continue to operate in the basis of cylindrical harmonics in order to facilitate comparison with the results and calculations of Section~\ref{BraidedQFT}, and also because we believe the perturbative calculations are clearer and more natural in this basis of states. Later on we will compare to the analogous calculations in the more traditional plane wave basis.

In particular, we use the same homotopy equivalence \eqref{SDR1} as before, which now leads to a quasi-isomorphism between the cochain complexes $(V,\mu_1)$ and $(H^\bullet(V),0)$ in the category $\CCV$. Since we shall again calculate vacuum correlation functions, we restrict to the trivial projection map $\sfp=0$. The homotopy $\sfh:V^2\longrightarrow V^1$ is then given by $\sfh = -\sgreen$, i.e. by the same formal expression (\ref{eq:htwo}) using the  Feynman propagator (\ref{TildeG}).

\subsection{Quantum correlation functions}

Lifting the $L_\infty$-structure \eqref{unbraidedell} to the strictly graded commutative algebra of functionals $\Sym (V[2])$ on $V$ no longer involves the non-trivial braiding $\RR$ and we can proceed in a straightforward fashion by directly applying the formulas from Section~\ref{QFT}.
Then the interacting quantum field theory is defined by the perturbation 
\begin{align*}
\mbf\delta^{\tt cyl} = -\ii\,\hbar\,\BVL - \{\CS^{\tt cyl} _{\rm int},-\} \ .
\end{align*}
The correlation functions are now given by 
\begin{equation}\label{eq:intcorrelationfnMomentum_2}
\widetilde C_n(p_1,\dots,p_n) = \sum_{m=1}^\infty \, \sP\,\big((-\ii\,\hbar\,\BVL\,\sH - \{\CS^{\tt cyl} _{\rm int},-\}\,\sH)^m\, (\ttc^{p_1}\odot\cdots\odot\ttc^{p_n})\big) \ , 
\end{equation}
 with the same basis of antifields $\ttc^p$ from \eqref{BasisAntiFields} as external states.
In the case of the free quantum field theory, i.e.~$\CS^{\tt cyl}_{\rm int}=0$ in (\ref{eq:intcorrelationfnMomentum_2}), this results in the standard (unbraided) Wick Theorem.

\subsection{Interacting Batalin-Vilkovisky master action}%

The interaction  functional $\CS^{\tt cyl} _{\rm int}\in \Sym(V[2])$ is defined by
\begin{equation}\label{eq:Sintzeta}
\CS^{\tt cyl} _{\rm int} := -\tfrac1{3!} \, \langle \xi \,,\,\mu^{{\rm ext}}_2(\xi,\xi)\rangle^{\rm ext} \ .
\end{equation}
The contracted coordinate functions are again given by the same formula
\begin{align*}
\xi = \int_{k}\hspace{-4.5mm}\mbox{$\sum$} \ \ \big(\ttc_k\otimes \ttc^k + \ttc^k\otimes \ttc_k\big) \ ,
\end{align*}
but now regarded as degree one elements of $\Sym(V[2])\otimes V$.
The computation of \eqref{eq:Sintzeta} is formally the same as the computation in Section~\ref{sub:intBVbraided} except that there are no $R$-matrices appearing now, and the only noncommutative contributions come from the explicit star-products appearing in (\ref{unbraidedell}). 

Symmetrization in the algebra of classical observables $\Sym (V[2])$, together with the fact that the binary bracket $\mu_2$ from \eqref{unbraidedell} is symmetric, results in
\begin{align}\label{SintFinal_2}
\CS^{\tt cyl}_{\rm int} = \int_{k_1}\hspace{-6mm}\mbox{$\sum$} \ \ \int_{k_2}\hspace{-6mm}\mbox{$\sum$} \ \ \int_{k_3}\hspace{-6mm}\mbox{$\sum$} \ \  V_{\tt cyl}(k_1, k_2, k_3) \ \ttc^{k_1}\odot \ttc^{k_2} \odot \ttc^{k_3}  \ ,
\end{align}
with the vertex 
\begin{align}\label{vertex2_2}
\begin{split}
V_{\tt cyl}(k_1, k_2, k_3) &= -\frac{g}{3!} \  \e^{-\frac{\ii\, \lambda}{2}\, \sum\limits_{a<b}\, (k_{az}\, \ell_{b} - k_{bz}\, \ell_{a})} \ (2\pi)^3 \ \delta(E_1+E_2+E_3) \ \delta(k_{1z}+k_{2z}+k_{3z})
 \\
&\hspace{7cm} \times  \delta_{\ell_{1} + \ell_{2} + \ell_{3}, 0} \ \ttF_{\ell_{1}, \ell_{2}, \ell_{3}} (\alpha_{1}, \alpha_{2}, \alpha_{3})  \ .
\end{split}
\end{align}
Up to the signs of the phase factors, the vertex \eqref{vertex2_2} is identical to the vertex \eqref{vertex2} of the braided theory. The difference in signs owes to the absence of $R$-matrix contributions in the present case.

\subsection{Two-point function at one-loop}%
\label{sub:2point1loop}

Let us now analyse in detail the two-point function $\widetilde C_2(p_1,p_2)$ at one-loop order following from (\ref{eq:intcorrelationfnMomentum_2}) and the analogous calculations of Section~\ref{BraidedQFT}.

\subsubsection*{Tree-level contribution}

The tree-level contribution is given by
\begin{align}
\begin{split}
\widetilde C_2(p_1, p_2)^{\swzero} &= -\ii\,\hbar\,\BVL\, \sH \, (\ttc^{p_1} \odot \ttc^{p_2})\\[4pt]
&= \ii\,\hbar\,(2\pi)^3\, (-1)^{\ell_{1}} \ \delta(E_1 + E_2) \ \delta(p_{1z} + p_{2z}) \ \delta_{\ell_{1} + \ell_{2}, 0} \ \frac{\delta(\alpha_{1} - \alpha_{2})}{\alpha_{1}} \ {\widetilde{\sgreen}}(p_1) \ .  \label{G2TreeLevel_2}
\end{split}
\end{align}
As expected, since the braided and standard free theories are the same, the tree-level two-point function \eqref{G2TreeLevel_2} is identical to (\ref{G2TreeLevel}).

\subsubsection*{One-loop corrections}

The one-loop contribution to the two-point function follows from 
\begin{align}
\begin{split}
\widetilde C_2(p_1, p_2)^{\swone} &=  \ii\,\hbar\,\BVL\,\sH \,\bigl\{ \CS^{\tt cyl} _{\rm int},\sH\, \big(\ii\,\hbar\,\BVL\,\sH\,\bigl\{\CS^{\tt cyl} _{\rm int},\sH\,(\ttc^{p_1}\odot \ttc^{p_2})\bigr\} \big)\bigr\} \\[4pt]
& \quad \, + (\ii\,\hbar\,\BVL\,\sH)^2 \, \bigl\{ \CS^{\tt cyl} _{\rm int},\sH\, \bigl\{\CS^{\tt cyl} _{\rm int},\sH\,(\ttc^{p_1}\odot \ttc^{p_2})\bigr\} \bigr\} \ .\label{TwoPoint_2}
\end{split}
\end{align}
The two terms in (\ref{TwoPoint_2}) include all possible contractions between the external momenta $(p_1,p_2)$ and internal momenta coming from a pair of vertices $(k_1, k_2, k_3)$ and $(q_1, q_2, q_3)$. Keeping in mind that we work in the graded commutative algebra $\Sym(V[2])$, i.e. there are no $R$-matrices that appear when elements in a symmetrized product are permuted, the standard Wick Theorem applies.

Let us summarise the different types of diagrammatic contributions  to $\widetilde C_2(p_1, p_2)^{\swone}$:
\begin{myitemize}
\item Contraction of the external momenta 
\begin{equation}
\BVL\, \sH \, (\ttc^{p_1} \odot \ttc^{p_2})   \nn
\end{equation}
results in  disconnected diagrams that consist of the free propagator diagram and two different vacuum diagrams:
\begin{equation}
{ \footnotesize \begin{tikzpicture}[scale=0.5, baseline]
        \coordinate[label=above:$p_1$] (p1) at (180:2.5);
        \coordinate[label=above:$p_2$] (p2) at (0:2.5);
        \draw[decoration={markings}, postaction={decorate}] (p1) -- (p2);
\end{tikzpicture} } \normalsize \quad \times  \quad \left( \ 
{ \footnotesize \begin{tikzpicture}[scale=0.5, baseline]
        \coordinate (k) at (-1,0);
        \coordinate (l) at (1,0);
        \draw[decoration={markings}, postaction={decorate}] ($(k)+(0:1)$) circle (0.5);  
        \draw[decoration={markings}, postaction={decorate}] ($(k) +(0:1) +(90:0.5)$) -- ($(k) +(0:1) +(270:0.5)$);
\end{tikzpicture} } \normalsize \qquad \text{or} \qquad
{ \footnotesize \begin{tikzpicture}[scale=0.5, baseline]
        \draw[decoration={markings}, postaction={decorate}] (0:2) circle (0.5);
        \draw[decoration={markings}, postaction={decorate}] (180:2) circle (0.5);
        \draw[decoration={markings}, postaction={decorate}] (0:1.5) -- (180:1.5);
\end{tikzpicture} } \normalsize \ \right)  \label{Disconnected1_2}
\end{equation}

\item Contractions between momenta from the same vertex
\begin{equation}
\BVL\, \sH \, (\ttc^{k_1} \odot \ttc^{k_2})  \nn
\end{equation}
 result in one reducible diagram and two disconnected diagrams of the form
\begin{equation}
{ \footnotesize \begin{tikzpicture}[scale=0.5, baseline]
        \coordinate (k) at (0,0);
        \coordinate (l) at ($(k)+(90:1)$);
        \coordinate[label=above: $p_1$] (p1) at ($(k) +(180:2.5)$);
        \coordinate[label=above: $p_2$] (p2) at ($(k) +(0:2.5)$);
\draw[decoration={markings, mark=at position 0.5 with {\arrow{Latex[reversed]}} }, postaction={decorate}] (k) -- (p1);
        \draw[decoration={markings, mark=at position 0.5 with {\arrow{Latex[reversed]}} }, postaction={decorate}] (k) -- (p2);
        \draw[decoration={markings}, postaction={decorate}] (k) -- (l);
        \draw[decoration={markings}, postaction={decorate}] ($(l)+(90:0.5)$) circle (0.5);
    \end{tikzpicture} } \normalsize \qquad \text{or} \qquad
 { \footnotesize    \begin{tikzpicture}[scale=0.5, baseline]
        \coordinate (k) at (-1.5,0);
        \coordinate (l) at (1.0,0);
        \coordinate[label=above: $p_1$] (p1) at (180:2.5);
        \coordinate[label=above: $p_2$] (p2) at (0:2.0);
\draw[decoration={markings, mark=at position 0.5 with {\arrow{Latex[reversed]}} }, postaction={decorate}] (k) -- (p1);
        \draw[decoration={markings, mark=at position 0.5 with {\arrow{Latex[reversed]}} }, postaction={decorate}] (l) -- (p2);
        \draw[decoration={markings}, postaction={decorate}] ($(k)+(0:0.5)$) circle (0.5);
        \draw[decoration={markings}, postaction={decorate}] ($(l)-(0:0.5)$) circle (0.5);
    \end{tikzpicture} } \normalsize \qquad \text{or} \qquad 
{ \footnotesize \begin{tikzpicture}[scale=0.5, baseline]
        \coordinate[label=above:$p_1$] (p1) at ($(0,0.5)+(180:2.5)$);
        \coordinate[label=above:$p_2$] (p2) at ($(0,0.5)+(0:2.5)$);
        \draw[decoration={markings}, postaction={decorate}] (p1) -- (p2);
        \draw[decoration={markings}, postaction={decorate}] ($(0,-0.5)+(0:2)$) circle (0.5);
        \draw[decoration={markings}, postaction={decorate}] ($(0,-0.5)+(180:2)$) circle (0.5);
        \draw[decoration={markings}, postaction={decorate}] ($(0,-0.5)+(0:1.5)$) -- ($(0,-0.5)+(180:1.5)$);
\end{tikzpicture} } \normalsize  \label{Disconnected2_2}
\end{equation}

\item Mixed contractions without permutations of momenta in the vertices
\begin{equation}
\BVL\, \sH \, (\ttc^{p_1} \odot \ttc^{k_1}) \ \BVL\, \sH \, (\ttc^{p_2} \odot \ttc^{q_1}) \ \BVL\, \sH \, (\ttc^{k_2} \odot \ttc^{q_3}) \ \BVL\, \sH \, (\ttc^{k_3} \odot \ttc^{q_2}) \label{ConnectedPlanar_2}
\end{equation}
 result in the planar one-loop contribution to the one-particle irreducible two-point function
\begin{equation}
{ \footnotesize \begin{tikzpicture}[scale=0.5, baseline]
        \coordinate (k) at (-0.5,0);
        \coordinate (l) at (0.5,0);
        \draw[decoration={markings, mark=at position 0.5 with {\arrow{Latex[reversed]}} }, postaction={decorate}] (k) -- ($(k) + (180:2)$) node[above]{$p_1$};
        \draw[decoration={markings, mark=at position 0.5 with {\arrow{Latex[reversed]}} }, postaction={decorate}] (l) -- ($(l) + (0:2)$) node[above]{$p_2$};
        \draw[decoration={markings}, postaction={decorate}] ($(k)+(0:0.5)$) circle (0.5);    
    \end{tikzpicture} } \normalsize \nn
\end{equation}

\item Mixed contractions with a single transposition of momenta in one vertex (e.g. $(q_1, q_2, q_3)\to (q_1, q_3, q_2)$)
\begin{equation}
\BVL\, \sH \, (\ttc^{p_1} \odot \ttc^{k_1}) \ \BVL\, \sH \, (\ttc^{p_2} \odot \ttc^{q_1}) \ \BVL\, \sH \, (\ttc^{k_2} \odot \ttc^{q_2}) \ \BVL\, \sH \, (\ttc^{k_3} \odot \ttc^{q_3}) \label{ConnectedNonPlanar_2}
\end{equation}
 introduce a single (trivial) braiding, resulting in the non-planar one-loop contribution to the one-particle irreducible two-point function
\begin{equation}
{ \footnotesize \begin{tikzpicture}[scale=0.5, baseline]
        \coordinate (k) at (-1,0);
        \coordinate (l) at (1,0);
        \draw[decoration={markings, mark=at position 0.5 with {\arrow{Latex[reversed]}} }, postaction={decorate}] (k) -- ($(k) + (180:2)$) node[above]{$p_1$};
        \draw[decoration={markings, mark=at position 0.5 with {\arrow{Latex[reversed]}} }, postaction={decorate}] (l) -- ($(l) + (0:2)$) node[above]{$p_2$};
        \draw[decoration={markings}, postaction={decorate}] (k) to[in=-90, out=-90, looseness=3] (0,0) to[in=90, out=90, looseness=3] (l);
        \draw[decoration={markings}, postaction={decorate}] (k) to[in=135, out=75, looseness=1.5] (-0.1,0.1);
        \draw[decoration={markings}, postaction={decorate}] (0.1,-0.1) to[in=-105, out=-45, looseness=1.5] (l);
    \end{tikzpicture} } \normalsize \nn
\end{equation}
\end{myitemize}

We discard the disconnected and reducible diagrams (\ref{Disconnected1_2}) and (\ref{Disconnected2_2}) as before, focusing only on the planar and non-planar contributions to the one-particle irreducible two-point function at one-loop. 

\paragraph{\underline{\sf Planar contribution.}}

In the notation of Section~\ref{sub:1loop2ptbraided}, the explicit form of a typical contribution from the planar diagram is
\begin{align}\label{G2FeynmanAdBPlanar_2}
\begin{split}
& \hspace{-1mm} \widetilde C_2(p_1, p_2)^{\swone}_{\text{planar}} \\[4pt] 
& \hspace{-3mm} \quad = \frac{1}{3} \, \frac{1}{2} \, (\ii\,\hbar)^2\, \Big(-\frac{g}{3!}\Big)^2\, (-1)^{\ell_{2}}\, (2\pi)^3 \ \delta(E_{1}+E_{2}) \ \delta(p_{1z}+p_{2z}) \ \delta_{\ell_{1}+\ell_{2}, 0} \ \widetilde{\sgreen}(p_1) \ \widetilde{\sgreen}(p_2)\\
& \hspace{-3mm}\hspace{1cm} \times  \ \int_{q_1}\hspace{-5.8mm}\mbox{$\sum$} \ \  \widetilde{\sgreen}(q_1) \ \int_0^\infty\, \alpha_{q_2} \, \d \alpha_{q_2} \ \frac{\ttF_{\ell_{1} + \ell_{q_1},-\ell_{q_1},-\ell_{1}} (\alpha_{q_2}, \alpha_{q_1}, \alpha_{1}) \, \ttF_{\ell_{q_1},-\ell_{2},-\ell_{1}-\ell_{q_1}} (\alpha_{q_1}, \alpha_{2}, \alpha_{q_2}) }{\alpha^2_{q_2} - (E_{1} + E_{q_1})^2 + (p_{1z} + q_{1z})^2 + m_+^2} \ .
\end{split}
\end{align}
Up to a symmetry factor, this diagram gives exactly the same (commutative) result as the one-particle irreducible diagram obtained from the braided BV formalism in (\ref{IIiIII}). In particular, it is logarithmically ultraviolet divergent.

\paragraph{\underline{\sf Non-planar contribution.}}

A typical contribution from the non-planar diagram is 
\begin{align} \label{G2FeynmanAdBNonPlanar_2}
\begin{split}
& \widetilde C_2(p_1, p_2)^{\swone}_{\text{non-planar}} \\[4pt] 
&  \quad = \frac{1}{3}\,\frac12 \, (\ii\,\hbar)^2\, \Big(-\frac{g}{3!}\Big)^2\, (-1)^{\ell_{2}}\, (2\pi)^3 \ \delta(E_{1}+E_{2}) \ \delta(p_{1z}+p_{2z}) \ \delta_{\ell_{1}+\ell_{2}, 0} \ \widetilde{\sgreen}(p_1) \ \widetilde{\sgreen}(p_2)\\
& \hspace{1cm} \times   \ \int_{q_1}\hspace{-5.8mm}\mbox{$\sum$} \ \ \e^{\,-\ii\,\lambda \, (p_{1z}\,\ell_{q_1} - q_{1z}\,\ell_1)} \ \widetilde{\sgreen}(q_1) \\
& \hspace{2cm} \times \ \int_0^\infty\, \alpha_{q_2} \, \d \alpha_{q_2} \ \frac{\ttF_{\ell_{1} + \ell_{q_1},-\ell_{q_1},-\ell_{1}} (\alpha_{q_2}, \alpha_{q_1}, \alpha_{1}) \, \ttF_{\ell_{q_1},-\ell_{2},-\ell_{1}-\ell_{q_1}} (\alpha_{q_1}, \alpha_{2}, \alpha_{q_2}) }{\alpha^2_{q_2} - (E_{1} + E_{q_1})^2 + (p_{1z} + q_{1z})^2 + m_+^2} \ .
\end{split}
\end{align}
This diagram leads to an obvious noncommutative contribution in the form of a phase factor mixing the internal and external momenta; in the braided theory, this factor is canceled by non-trivial $R$-matrix contributions from the Braided Wick Theorem, effectively reducing the non-planar diagram to the planar diagram. As expected, in the $\lambda=0$ theory the contribution (\ref{G2FeynmanAdBNonPlanar_2}) reduces to the planar contribution (\ref{G2FeynmanAdBPlanar_2}) and hence to the commutative result  (up to a symmetry factor). 

\paragraph{\underline{\sf Periodic UV/IR mixing.}}

Let us now explore the limit of the two-point function \eqref{G2FeynmanAdBNonPlanar_2} in which the phase factor tends to unity when $\lambda\neq0$. In the cylindrical harmonic basis, this occurs as the limit of external momentum in which the angular momentum vanishes, $\ell_{1} \to 0$, and the axial momentum becomes \emph{exceptional}, i.e. $p_{1z}$ approaches one of the infinitely many isolated points with $\lambda\,p_{1z}\in 2\pi\,\RZ$ and $\sin(\frac\lambda2\,p_{1z})\to0$. In this `periodic' infrared limit, exceptional momenta render the phase factor ineffective and the whole non-planar correlator (\ref{G2FeynmanAdBNonPlanar_2}) reduces to the planar correlator \eqref{G2FeynmanAdBPlanar_2}. This is a potential signal of a periodic form of UV/IR mixing, in that the original ultraviolet divergence reappears as a periodic infrared divergence.

To substantiate this claim, as in Section~\ref{sub:1loop2ptbraided} we focus on the special case where the momentum is purely axial, that is, $\alpha_1=0$ and $\lambda\,p_{1z}\notin 2\pi\,\RZ$. The triple Bessel integrals are completely symmetric under any permutation of pairs $(\alpha_i, \ell_i)$ and we can sum over $\ell_{q_1} \rightarrow -\ell_{q_1}$. 
From (\ref{Jzerozero}) it is clear that the phase factor in \eqref{G2FeynmanAdBNonPlanar_2} will not affect the Feynman parametrisation introduced earlier in (\ref{UVStep1}) and we can write
\begin{align}\label{Pomocna1}
\begin{split}
& \widetilde C_2(p_1, p_2)^{\swone}_{\text{non-planar}} \, \Big|_{\alpha_1=0} \\[4pt] 
& \qquad =\frac{\ii\, \pi}{6} \, (\ii\,\hbar)^2 \, \Big(-\frac{g}{3!}\Big)^2 \ \delta(E_{1}+E_{2}) \ \delta(p_{1z}+p_{2z}) \ \delta_{\ell_{1}, 0} \ \delta_{\ell_{2}, 0} \ \widetilde{\sgreen}(p_1) \ \widetilde{\sgreen}(p_2)\\
& \hspace{2cm} \times \sum_{\ell_{q_1}\in\RZ} \ \int_{0^+}^\infty\, \alpha_{q_1} \, \d \alpha_{q_1} \
\e^{\,\ii\,\lambda \, p_{1z}\,\ell_{q_1}} \ \int_0^1 {\rm d}x \ \frac{(-1)^{\ell_{q_1}} \, \ttF_{\ell_{q_1},0,-\ell_{q_1}} (\alpha_{q_1}, \alpha_{2}, \alpha_{q_1}) }{\alpha^2_{q_1} + m_+^2-x\,(1-x)\,(E_1^2-p_{1z}^2)}   \ .
\end{split}
\end{align}

The second triple Bessel integral in \eqref{Pomocna1} is given by
\begin{equation}
\ttF_{\ell_{q_1},0,-\ell_{q_1}} (\alpha_{q_1}, \alpha_{2}, \alpha_{q_1}) = (-1)^{\ell_{q_1}}\,\frac{\cos(\ell_{q_1}\,\upsilon_2)}{\pi\, \alpha_{q_1}\,\alpha_2\cos (\frac{\upsilon_2}{2})} \ , \nn
\end{equation}
and the corresponding momentum triangle is isosceles. The sum over $\ell_{q_1}\in\RZ$ results in a pair of Dirac distributions
\begin{align}
\sum_{\ell_{q_1}\in\RZ}\, \e^{\,\ii\,\lambda \, p_{1z}\,\ell_{q_1}}  \cos(\ell_{q_1}\upsilon_2) =  \pi\,\delta(\upsilon_2\,+\lambda \, p_{1z}) + \pi\,\delta(\upsilon_2\,-\lambda \, p_{1z}) \ .\nn
\end{align}
As in Section~\ref{sub:1loop2ptbraided}, the two Dirac distributions cannot be simultaneously non-zero, and which one can be discarded depends on the sign of $\lambda\,p_{1z}$. The non-vanishing  distribution in the angles can then be expressed as a  distribution in the sides of the triangle with the aid of the triangle constraint
\begin{equation}\nn
\sin\Big(\frac{\upsilon_2}{2}\Big) = \frac{\alpha_2}{2\,\alpha_{q_1}} \ .
\end{equation}

To do this, we treat $\delta(\sin(\frac12\,(\upsilon_2\mp\lambda\, p_{1z}))$ as a function of $\upsilon_2$ and use $\delta(\sin(\upsilon))=\delta(\upsilon)$ for $\upsilon\in[0,\pi)$ to get
\begin{equation}
\delta\Big(\sin\big(\tfrac12\,(\upsilon_2\mp\lambda\, p_{1z})\big)\Big) = \delta\big(\tfrac12\,(\upsilon_2\mp\lambda\, p_{1z})\big) = 2\, \delta(\upsilon_2\mp\lambda\, p_{1z}) \ .\label{Pomocna2}
\end{equation}
On the other hand, $\delta(\sin(\frac12\,(\upsilon_2\mp\lambda\, p_{1z}))$ can also be understood as a function of $\alpha_2$, which yields
\begin{align}
\begin{split}
\delta\Big(\sin\big(\tfrac12\,(\upsilon_2\mp\lambda\, p_{1z})\big)\Big) &= \delta\big(\sin(\tfrac12\,\upsilon_2)\cos(\tfrac{\lambda}{2}\,p_{1z}) \mp \cos(\tfrac12\,\upsilon_2)\sin(\tfrac{\lambda}{2}\,p_{1z})\big) \\[4pt]
&=\delta\bigg(\frac{\alpha_2}{2\,\alpha_{q_1}}\cos\big(\tfrac{\lambda}{2}\,p_{1z}\big)\mp\sqrt{1-\Big(\frac{\alpha_2}{2\,\alpha_{q_1}}\Big)^2}\,\sin\big(\tfrac{\lambda}{2}\,p_{1z}\big)\bigg)\label{Pomocna3}\\[4pt]
&= 2\, \alpha_{q_1}\cos(\tfrac{\lambda}{2}\, p_{1z})\,\delta\big(\alpha_2\mp 2\,\alpha_{q_1}\sin(\tfrac\lambda2\, p_{1z})\big) \ . 
\end{split}
\end{align}
Equating (\ref{Pomocna2}) and  (\ref{Pomocna3}) we find
\begin{equation}
\delta(\upsilon_2\mp\lambda\, p_{1z})= \alpha_{q_1}\cos\big(\tfrac{\lambda}{2}\,p_{1z}\big)\,\delta\big(\alpha_2\mp2\,\alpha_{q_1}\sin(\tfrac\lambda2\, p_{1z})\big) \ , \nn
\end{equation}
where the sign $\mp$ corresponds respectively to ${\rm sgn}(\lambda\,p_{1z})=\pm 1$.

Inserting this result into (\ref{Pomocna1}) gives
\begin{align}
\begin{split}
& \widetilde C_2(p_1, p_2)^{\swone}_{\text{non-planar}} \, \Big|_{\alpha_1=0} \\[4pt] 
& \hspace{2cm} = \frac{\ii\, \pi}{6} \, (\ii\,\hbar)^2\, \Big(-\frac{g}{3!}\Big)^2 \ \delta(E_{1}+E_{2}) \ \delta(p_{1z}+p_{2z}) \ \delta_{\ell_{1}, 0} \ \delta_{\ell_{2}, 0} \ \widetilde{\sgreen}(p_1) \ \widetilde{\sgreen}(p_2) \\
& \hspace{4cm} \times \int_{0^+}^\infty\, \alpha_{q_1} \, \d \alpha_{q_1} \ \int_0^1\, {\rm d}x \ \frac{\delta\big(\alpha_2\mp 2\,\alpha_{q_1}\sin(\tfrac\lambda2\, p_{1z})\big) }{\alpha_2\,\big(\alpha^2_{q_1} + m_+^2-x\,(1-x)\,(E_1^2-p_{1z}^2)\big)}  \ .\label{UVfiniteCyl}
\end{split}
\end{align}
Performing the integration over $\alpha_{q_1}\in(0,\infty)$ and subsequently over $x\in[0,1]$ finally leads to
\begin{align} \label{UVFiniteCylAdd}
\begin{split}
& \widetilde C_2(p_1, p_2)^{\swone}_{\text{non-planar}} \, \Big|_{\alpha_1=0} \\[4pt] 
& \hspace{2cm} = \frac{\ii\, \pi}{12} \, (\ii\,\hbar)^2\, \Big(-\frac{g}{3!}\Big)^2 \ \delta(E_{1}+E_{2}) \ \delta(p_{1z}+p_{2z}) \ \delta_{\ell_{1}, 0} \ \delta_{\ell_{2}, 0} \ \widetilde{\sgreen}(p_1) \ \widetilde{\sgreen}(p_2)\\
& \hspace{3cm} \times \frac{\displaystyle\tan^{-1}\sqrt{\frac{\sin^2(\tfrac\lambda2\,p_{1z})\,(E_1^2-p_{1z}^2)}{\alpha_2^2+\sin^2(\tfrac\lambda2\,p_{1z})\,\big(4m^2-(E_1^2-p_{1z}^2)\big)}}}{\big|\sin\big(\tfrac\lambda2\,p_{1z}\big)\big|\,\sqrt{\big(E_1^2-p_{1z}^2\big)\,\big(\alpha_2^2+\sin^2(\tfrac\lambda2\,p_{1z})\,\big(4m^2-(E_1^2-p_{1z}^2)\big)\big)}} \ .
\end{split}
\end{align}

At any non-exceptional momentum $\lambda\,p_{1z}\notin 2\pi\,\RZ$, the non-planar contribution \eqref{UVFiniteCylAdd} is finite, in contrast to its logarithmically divergent planar counterpart, and the radial momentum $\alpha_2$ is no longer conserved thanks to the breaking of translational invariance by noncommutativity. However, taking the limit $\lambda\,p_{1z}\in 2\pi\,\RZ$ in (\ref{UVfiniteCyl}) removes the radial momentum $\alpha_{q_1}$ from the Dirac distribution and leaves $\alpha_2$ trivially conserved. Thus in either the commutative or periodic infrared limit we completely recover the planar contribution (\ref{UVStep4}) and its logarithmic ultraviolet divergence. Because of this non-analytic behaviour of \eqref{UVFiniteCylAdd} as a function of $\lambda\,p_{1z}$, we conclude that periodic UV/IR mixing is present in the standard BV-quantized $\varPhi^3_\star$-theory represented in the basis of cylindrical harmonic states. 
 
\subsection{Correlation functions in the plane wave basis} \label{sub:planewave} %

As already discussed, in contrast to the braided BV formalism from Section~\ref{BraidedQFT}, in the standard BV formalism there are no representation theoretic restrictions needed to define the noncommutative quantum field theory. From this perspective, instead of the cylindrical harmonic basis (\ref{BasisFields}) and (\ref{BasisAntiFields}) of fields and antifields, we could just as well appeal to the more conventional Poincar\'e invariant basis \eqref{eq:planewaves} of plane waves 
\begin{align}
\tte_k (x) &:= \tte_{(E,k_x,k_y,k_z)}(t,x,y,z)  \ \in \ V^1 \ ,\label{BasisFields_2}\\[4pt]
\tte^k (x) &:= \tte_{-k}(x) = \tte_{(-E,-k_x,-k_y,-k_z)}(t,x,y,z) \ \in \ V^2 \ , \label{BasisAntiFields_2}
\end{align}
defined via four-vectors $x$ in position space and standard four-momenta $k$. These basis vectors are indeed dual with respect to the cyclic structure \eqref{CyclPairing_2}:
\begin{align}\nn
    \langle \tte_{k_1}\,,\,\tte^{k_2}\rangle = \int_{\FR^4}\, \dd^4x \ \e^{-\ii\,(k_1-k_2)\cdot x} = (2\pi)^4 \ \delta(k_1-k_2) \ .
\end{align}

Let us now discuss the standard BV formalism for $\varPhi_\star^3$-theory in this basis, and compare the results with the calculations performed thus far in the cylindrical harmonic basis. There are two places where the change of basis is relevant: 
\begin{myitemize}
\item The interacting BV master action $\CS^{\tt plwa} _{\rm int}\in \Sym(V[2])$ from \eqref{S_celo_BV} in the plane wave basis is  defined by
\begin{equation}\label{eq:S_BV_plwa}
\CS^{\tt plwa} _{\rm int} := -\tfrac1{3!} \, \langle \zeta \,,\,\mu^{{\rm ext}}_2(\zeta,\zeta)\rangle^{\rm ext} \ ,
\end{equation}
where now the contracted coordinate functions $\zeta\in\Sym(V[2])\otimes V$ are constructed as combinations of plane waves
\begin{align*}
\zeta = \int_{\FR^4}\, \frac{\dd^4 k}{(2\pi)^4} \ \big(\tte_k\otimes \tte^k + \tte^k\otimes \tte_k\big) \ .
\end{align*}
\item Homological perturbation theory now constructs the $n$-point quantum correlation functions \eqref{eq:BQFTnpoint} in momentum space as
\begin{equation}\label{eq:tildeGn}
\widetilde G_n(p_1,\dots,p_n) = \sum_{m=1}^\infty \, \sP\,\big((-\ii\,\hbar\,\BVL\,\sH - \{\CS^{\tt plwa} _{\rm int},-\}\,\sH)^m\, (\tte^{p_1}\odot\cdots\odot\tte^{p_n})\big) \ , 
\end{equation}
with  insertions of the plane wave basis of antifields $\tte^p$ from \eqref{BasisAntiFields_2} as external states.
\end{myitemize}

To describe these modifications explicitly, following~\cite{DimitrijevicCiric:2018blz} we deform the addition law of momenta according to the deformed coproducts of the Cartesian components of the momentum operator in (\ref{TwistedPoincareCoproduct}). We define the noncommutative associative star-sum $k+_\star p$ of momenta $k,p\in\FR^4$ through $$\tte_{k+_\star p} := \tte_k\star\tte_p  \ . $$ 
This binary operation deforms the additive abelian group $(\FR^4,+)$ to the non-abelian group $(\FR^4,+_\star)$, with unit $0$ and where the inverse of $k$ with respect to $+_\star$ is again $-k$.

Explicitly, the star-sums of two and three momenta read as~\cite{DimitrijevicCiric:2018blz}
\begin{align}
k+_\star p &= \sfLambda(p_z)\,k + \sfLambda(-k_z)\,p  \ ,\label{StarSumTwo_2}\\[4pt]
k +_{\star} p +_{\star} q &= \sfLambda(p_z + q_z)\,k + \sfLambda(-k_z + q_z)\,p  + \sfLambda(-k_z -p_z)\,q \ , \label{StarSumThree_2}
\end{align}
where the rotation matrix
\begin{equation}
\sfLambda(p_z) = {\small \left( \begin{matrix}
     1 & 0 & 0 & 0\\
     0 & \cos \big(\frac{\lambda}{2}\,p_z\big) & \sin \big(\frac{\lambda}{2}\,p_z\big) &0\\
     0 & -\sin \big(\frac{\lambda}{2}\,p_z\big) & \cos \big(\frac{\lambda}{2}\,p_z\big) &0\\
     0 & 0 & 0 & 1
    \end{matrix} \right) } \normalsize \nn
\end{equation}
implements a rotation through angle $\frac\lambda2\,p_z$ in the $(x,y)$-plane of momentum space. While the star-sum does not affect the components $(E,k_z)$ (they are summed in the usual abelian way), it modifies the addition law of the components $(k_x,k_y)$.

The deformed momentum conservation laws implied by \eqref{TwistedPoincareCoproduct} are then implemented by the Dirac distribution
\begin{align}
\delta (k_1 +_\star \cdots +_\star k_n) =  \int_{\FR^4}\, \frac{\d^4x}{(2\pi)^4} \ \e^{\,\ii\,k_1\cdot x}\star \cdots \star \e^{\,\ii\,k_n\cdot x} \ .\label{StarDelta3_2} 
\end{align}
By the cyclicity property \eqref{IntegralCyc}, it satisfies
\begin{align}\nn
\delta (k_1 +_\star k_2 +_\star \cdots +_\star k_n)=\delta (k_2 +_\star \cdots +_\star k_n +_\star k_1)
= \delta \big(k_1 +  (k_2 +_\star \cdots +_\star k_n)\big) \ .
\end{align}
In particular, $\delta(k_1+_\star k_2) = \delta(k_1+k_2)$.

Calculating as previously thereby leads to the interaction functional \eqref{eq:S_BV_plwa} in the form
\begin{align}\nn
\CS^{\tt plwa}_{\rm int} = \int_{(\FR^{4})^{\times 3}}\, \frac{\dd^4k_1}{(2\pi)^4} \ \frac{\dd^4k_2}{(2\pi)^4} \ \frac{\dd^4k_3}{(2\pi)^{4}} \ V_{\tt plwa}(k_1, k_2, k_3) \ \tte^{k_1}\odot \tte^{k_2} \odot \tte^{k_3}  \ ,
\end{align}
where the vertex \eqref{vertex2_2} changes to a star-sum of momenta implemented by a Dirac distribution
\begin{align}\label{VertexFeynmanPlane_2}
\begin{split}
V_{\tt plwa}(k_1, k_2, k_3) &= -\frac{g}{3!} \,   (2\pi)^4 \ \delta(k_1+_\star k_2+_\star k_3) \ .
\end{split}
\end{align}
In particular, in this basis the noncommutativity of the interaction is no longer captured by a simple Moyal-like phase factor, but rather by a vertex which is formally identical to that of the commutative theory with a deformed momentum conservation law.

\subsubsection*{Two-point function at one-loop}

Performing identical calculations to before but now in the plane wave basis, the free scalar field theory gives rise to the two-point function
\begin{align}
\widetilde G_2(p_1, p_2)^{\swzero} = -\ii\,\hbar\,\BVL\, \sH \, (\tte^{p_1} \odot \tte^{p_2}) =  \ii\,\hbar \,(2\pi)^4 \ \delta(p_1 + p_2) \ {\widetilde{\sgreen}}(p_1) \ . \label{G2TreeLevel_2p}
\end{align}
The diagrams contributing to the one-loop corrections to \eqref{G2TreeLevel_2p} are the same as in the cylindrical harmonic basis, and again we just present the results of the calculations for the planar and non-planar contributions to the one-particle irreducible two-point function at one-loop.

\paragraph{\underline{\sf Planar contribution.}}

Starting from (\ref{ConnectedPlanar_2}), the typical planar diagram contribution to the one-loop two-point function is given by
\begin{align}
\begin{split}
\widetilde G_2(p_1, p_2)^{\swone}_{\text{planar}} &= \frac13\,\frac12\,(\ii\,\hbar)^2\,\Big(-\frac{g}{3!}\Big)^2 \, (2\pi)^4 \ \delta(p_1+p_2) \ \widetilde{\sgreen}(p_1) \ \widetilde{\sgreen}(p_2) \\
    &\hspace{6cm} \times \int_{\FR^4} \, \frac{\dd^4k}{(2\pi)^4} \ \widetilde{\sgreen}\big((-p_1) +_\star k\big) \, \widetilde{\sgreen}(k) \ . \label{G2FeynmanPlanarPlane_2}
    \end{split}
\end{align}
The first loop propagator contains an apparent noncommutative contribution in the form of a star-sum of external and internal momenta. A similar behaviour for the planar diagram contribution to the one-loop four-point function in the standard $\varPhi^4_\star$-theory was observed in~\cite{Hersent:2023lqm, Wallet:2025mbv}. 

Let us look more closely at (\ref{G2FeynmanPlanarPlane_2}). 
Using (\ref{StarSumTwo_2}) and converting to polar coordinates in the $(p_x,p_y)$-plane, with $p_{ix}=\alpha_i\cos(\vartheta_i)$ and $p_{iy}=\alpha_i\sin(\vartheta_i)$ (see Appendix~\ref{app:Bessel}), the Minkowski norm of the star-sum $(-p_1) +_\star k$ can be rewritten as
\begin{align*}
    \big((-p_1) +_\star k\big)^2 &= E_{(-p_1) +_\star k}^2 - \big((-p_1) +_\star k\big)^2_x- \big((-p_1) +_\star k\big)^2_y- \big((-p_1) +_\star k\big)^2_z \\[4pt]
    &= (E_k - E_1)^2 - (k_z - p_{1z})^2 - \big(\alpha^2_{1} + \alpha^2_{k}\big) + 2\, \alpha_{1}\,\alpha_{k} \cos\big(\vartheta_{k} - \vartheta_{1}- \tfrac{\lambda}{2}\,(p_{1z}-k_z)\big) \ .
\end{align*}
Written in cylindrical coordinates, the loop integral in (\ref{G2FeynmanPlanarPlane_2}) is then
\begin{align*}
&\int_{\FR^4} \, \frac{\dd^4k}{(2\pi)^4} \ \widetilde{\sgreen}\big((-p_1) +_\star k\big) \, \widetilde{\sgreen}(k) \\[4pt]
& \quad = \int_{\FR\times\FR} \, \frac{\dd E_k\,\dd k_z}{(2\pi)^2} \ \int_0^\infty \, \frac{\alpha_k\, \dd \alpha_k}{2\pi} \  \int_0^{2\pi} \, \frac{\dd \vartheta_k}{2\pi} \ \frac{1}{E^2_k - k^2_z - \alpha^2_k - m_+^2}\\
& \hspace{1cm} \times \frac{1}{(E_k - E_1)^2 - (k_z - p_{1z})^2 - \big(\alpha^2_{1} + \alpha^2_{k}\big) + 2\, \alpha_{1}\,\alpha_{k} \cos\big(\vartheta_{k} - \vartheta_{1}- \frac{\lambda}{2}\,(p_{1z}-k_z)\big) - m_+^2} \ .
\end{align*}

The integrand is periodic with respect to $\vartheta_k$, and the angular integral is over $\vartheta_k\in\FR/2\pi\,\RZ$. Therefore, since $\FR/2\pi\,\RZ$ is an $\FR$-module, the integral does not depend on arbitrary constant translations of $\vartheta_k$. If we choose to translate $\vartheta_k \rightarrow \vartheta_k + \frac{\lambda}{2}\,(p_{1z} - k_z)$, we eliminate the dependence on the deformation parameter $\lambda$ and restore the commutative result
\begin{align*}
\begin{split}
\widetilde G_2(p_1, p_2)^{\swone}_{\text{planar}} &= \frac16\,(\ii\,\hbar)^2\,\Big(-\frac{g}{3!}\Big)^2 \, (2\pi)^4 \ \delta(p_1+p_2) \ \widetilde{\sgreen}(p_1) \ \widetilde{\sgreen}(p_2) \ \int_{\FR^4} \, \frac{\dd^4k}{(2\pi)^4} \ \widetilde{\sgreen}\big(k-p_1\big) \, \widetilde{\sgreen}(k) \ . \end{split}
\end{align*}
In this way we showed that the planar contribution to the one-loop two-point function is the same as the commutative result (up to a symmetry factor). This is consistent with the planar equivalence theorem \cite{Meier:2023lku} and the result \eqref{G2FeynmanAdBPlanar_2}. 

\paragraph{\underline{\sf Non-planar contribution.}}

Starting from \eqref{ConnectedNonPlanar_2} the typical non-planar diagram contribution to the one-loop two-point function is given by
\begin{align}
\begin{split}
\widetilde G_2(p_1, p_2)^{\swone}_{\text{non-planar}} &= \frac13\,\frac12\,(\ii\,\hbar)^2\,\Big(-\frac{g}{3!}\Big)^2 \ \widetilde{\sgreen}(p_1) \ \widetilde{\sgreen}(p_2)\\
    & \hspace{1cm} \times \int_{\FR^4} \, \frac{\dd^4k}{(2\pi)^4}  \ \widetilde{\sgreen}\big((-p_1) +_\star k\big) \, \widetilde{\sgreen}(k) \\[-5pt]
    & \hspace{4cm} \times (2\pi)^4 \ \delta\big(-p_2 +_\star (-k) +_\star (-p_1) +_\star k \big) \ . \label{G2FeynmanNonPlanarPlane_2} 
    \end{split}
\end{align}
The non-planar correction \eqref{G2FeynmanNonPlanarPlane_2} receives two noncommutative contributions:
\begin{myitemize}
\item The first one comes from the Dirac distribution: it represents the deformed conservation law of momentum in the $(x,y)$-plane, i.e. of the radial momentum. This contribution has as its origin the star-product of plane waves, and is analogous to the phase factors appearing in the non-planar contribution to the corresponding correlator in the cylindrical harmonic basis \eqref{G2FeynmanAdBNonPlanar_2}.

\item The second one comes from the star-sum of momenta in a loop propagator and is specific to the $\lambda$-Minkowski (and $\rho$-Minkowski) space \cite{Hersent:2023lqm, Wallet:2025mbv}. 
\end{myitemize}

Following \cite{DimitrijevicCiric:2018blz}, one can check that the ultraviolet behaviour of (\ref{G2FeynmanNonPlanarPlane_2}) is indeed improved at any non-exceptional external momentum $\lambda\,p_{1z}\notin 2\pi\,\RZ$. The explicit form of the Dirac distribution
\begin{align}
\begin{split}
& \delta\big(-p_2 +_\star (-k) +_\star (-p_1) +_\star k \big) \\[4pt]
& \hspace{2cm} = \delta(E_{1} + E_{2}) \ \delta(p_{1z} + p_{2z}) \ \frac{1}{4\sin^2 \big(\frac{\lambda}{2}\,p_{2z} \big)} \label{UVfiniteCart}\\
& \hspace{3cm} \times \delta\Big( k_x - \frac{p_{1y} + p_{2y}}{2\sin \big(\frac{\lambda}{2}\,p_{2z} \big)}\cos \big(\tfrac{\lambda}{2}\,k_z \big) 
+ \frac{p_{1x} - p_{2x}}{2\sin \big(\frac{\lambda}{2}\,p_{2z} \big)}\sin \big(\tfrac{\lambda}{2}\,k_z \big) \Big) \\
& \hspace{4cm} \times \delta\Big( k_y + \frac{p_{1x} + p_{2x}}{2\sin \big(\frac{\lambda}{2}\,p_{2z} \big)}\cos \big(\tfrac{\lambda}{2}\,k_z \big) 
+ \frac{p_{1y} - p_{2y}}{2\sin \big(\frac{\lambda}{2}\,p_{2z} \big)}\sin \big(\tfrac{\lambda}{2}\,k_z \big) \Big) 
\end{split}
\end{align}
enables integration of the $(k_x,k_y)$ components of the loop momentum $k$ leaving an overall contribution of order $\int_{\FR\times\FR}\, \frac{\d E_k\, \d k_z}{k^4}$ in the ultraviolet limit, which converges. 

In order to compare the results (\ref{G2FeynmanNonPlanarPlane_2}) and (\ref{G2FeynmanAdBNonPlanar_2}) explicitly, we look more closely into the special case when $\alpha_1=0$. The deformation of the loop momentum can then be simplified due to the form of the deformed momentum conservation law (\ref{UVfiniteCart}) which gives
\begin{align*}
    \big((-p_1) +_\star k\big)^2\,\Big|_{\alpha_1=0} = (E_k - E_1)^2 - (k_z - p_{1z})^2 - \frac{\alpha^2_2}{4\sin^2\big(\frac{\lambda}{2}\,p_{2z}\big)} \ .
\end{align*}
The contribution of the non-planar diagram (\ref{G2FeynmanNonPlanarPlane_2}) then reduces to
\begin{align*}
\begin{split}
 \widetilde G_2(p_1, p_2)^{\swone}_{\text{non-planar}} \, \Big|_{\alpha_1=0} \normalsize &= \frac1{6} \, (\ii\,\hbar)^2\,\Big(-\frac{g}{3!}\Big)^2 \  \delta(E_1 + E_2) \ \delta(p_{1z} + p_{2z}) \ \frac{\widetilde{\sgreen}(p_1) \ \widetilde{\sgreen}(p_2)}{4\sin^2\big(\frac{\lambda}{2}\,p_{2z}\big)}\\
    & \quad \, \times \int_{\FR\times\FR} \, \d E_k\, \d k_z  \ \frac{1}{(E_k - E_1)^2 - (k_z - p_{1z})^2 - \frac{\alpha^2_2}{4\sin^2(\frac{\lambda}{2}\,p_{2z})} - m_+^2}\\
    & \hspace{5cm} \times \frac{1}{E^2_k - k^2_z - \frac{\alpha^2_2}{4\sin^2(\frac{\lambda}{2}\,p_{2z})} - m_+^2} \ .
    \end{split}
\end{align*}

The loop integral is now formally identical to (\ref{UVStep1}) and it can be evaluated in a similar manner. The end result is finite and after integration over the Feynman parameter $x\in[0,1]$ it is given by
\begin{align}
\begin{split}
& \widetilde G_2(p_1, p_2)^{\swone}_{\text{non-planar}} \, \Big|_{\alpha_1=0} \\[4pt]
& \hspace{2cm}= \frac{\ii\, \pi}{12} \, (\ii\,\hbar)^2\, \Big(-\frac{g}{3!}\Big)^2 \ \delta(E_{1}+E_{2}) \ \delta(p_{1z}+p_{2z}) \ \widetilde{\sgreen}(p_1) \ \widetilde{\sgreen}(p_2)\\
& \hspace{3cm} \times \frac{\displaystyle\tan^{-1}\sqrt{\frac{\sin^2(\frac\lambda2\,p_{1z})\,(E_1^2-p_{1z}^2)}{\alpha_2^2+\sin^2(\frac\lambda2\,p_{1z})\,\big(4m^2-(E_1^2-p_{1z}^2)\big)}}}{\big|\sin\big(\frac\lambda2\,p_{1z}\big)\big|\,\sqrt{\big(E_1^2-p_{1z}^2\big)\,\big(\alpha_2^2+\sin^2(\frac\lambda2\,p_{1z})\,\big(4m^2-(E_1^2-p_{1z}^2)\big)\big)}}  \ . \label{G2FeynmanNonPlanarPlane_2Add}
\end{split}
\end{align}

This is precisely the same result we found for the non-planar contribution in the cylindrical harmonic basis (\ref{UVFiniteCylAdd}). In particular, the noncommutative contribution to the non-planar correction again removes the logarithmic ultraviolet divergence and gives a finite result for all $\lambda\,p_{1z}\notin 2\pi\,\RZ$. Note that in both (\ref{UVFiniteCylAdd}) and (\ref{G2FeynmanNonPlanarPlane_2Add}) the external momentum is not conserved in the $(x,y)$-plane, due to the absence of Dirac distributions $\delta(\alpha_{1} - \alpha_{2})$ and $\delta(p_{1x} + p_{2x})\,\delta(p_{1y} + p_{2y})$ (equivalently $\delta(\alpha_1-\alpha_2)\,\delta(\vartheta_1-\vartheta_2-\pi)$) in (\ref{UVFiniteCylAdd}) and (\ref{G2FeynmanNonPlanarPlane_2Add}) correspondingly, reflecting the breaking of translational invariance by the noncommutative deformation.

On the other hand, in the limit $\alpha_1\to0$ and $\lambda\,p_{1z}\in 2\pi\,\RZ$ the contribution (\ref{G2FeynmanNonPlanarPlane_2}) reduces to the planar contribution (\ref{G2FeynmanPlanarPlane_2}). As in the planar diagram contribution, the logarithmic ultraviolet divergence reappears in the periodic infrared limit, and therefore we confirm that periodic UV/IR mixing is also present in the standard BV-quantized $\varPhi^3_\star$-theory represented in the basis of plane wave states. 

The results (\ref{UVFiniteCylAdd}) and (\ref{G2FeynmanNonPlanarPlane_2Add}) suggest that periodic UV/IR mixing is a consequence of the standard BV quantization and that it is a basis independent phenomenon.  It appears as a consequence of the angular twist deformation and the  breaking of  translational invariance in the $(x,y)$-plane. Compared to the more standard form of the UV/IR mixing in the Moyal case, periodic UV/IR mixing is far more pathological: as the axial momentum $p_z\in\FR$ is not a periodic coordinate, in higher-loop integrals infinitely many exceptional momenta are attained and contribute uncontrollable infinities. In particular, it is not clear how the usual completions via the addition of a background magnetic field~\cite{Langmann:2002cc} or a harmonic oscillator potential~\cite{Grosse:2004yu}, which provide  infrared regulators eliminating the usual form of UV/IR mixing, can be extended to this periodic case.

\subsection{Plane waves versus cylindrical harmonics}
\label{PWversusCY}%

Our results for the two-point function at one-loop order in the two different bases, the planar (\ref{G2FeynmanAdBPlanar_2}) and non-planar (\ref{G2FeynmanAdBNonPlanar_2}) diagram contributions in the basis of cylindrical harmonic states, and the planar (\ref{G2FeynmanPlanarPlane_2}) and non-planar (\ref{G2FeynmanNonPlanarPlane_2}) diagram contributions in basis of plane wave states, agree precisely. The planar diagrams are logarithmically ultraviolet divergent, just as in the commutative case. The non-planar diagrams are ultraviolet finite, but in the limit $\lambda\,p_{1z}\in 2\pi\,\RZ$ they reduce to the corresponding planar diagrams and the ultraviolet divergence reappears, resulting in periodic UV/IR mixing. To better understand the relation between these results, we shall now compare the computations in these two different bases. We will decompose our building blocks --- the vertex (\ref{VertexFeynmanPlane_2}) and contractions, i.e. the free propagator (\ref{G2TreeLevel_2p}) --- given in terms of plane waves into the corresponding objects given in terms of cylindrical harmonics.

Recall that plane waves are characterized by quadruples of momenta $(E_k, k_x, k_y, k_z)$ in Cartesian coordinates, or equivalently by quadruples of momenta $(E_k, \alpha_k, \vartheta_k, k_z)$ in cylindrical coordinates, where 
\begin{align} \label{eq:radialpolartransf}
\alpha_k = \sqrt{k^2_x+k^2_y} \qquad , \qquad \vartheta_k =  \tan^{-1}\Big(\frac{k_y}{k_x}\Big)
\end{align}
are the radial momentum and the polar angle respectively in the $(k_x,k_y)$-plane. Cylindrical harmonics are instead characterised by quadruples of momenta $(E_k, \alpha_k, \ell_k, k_z)$, where $\ell_k\in\RZ$ is the angular momentum in the $(k_x,k_y)$-plane.

Plane waves in Cartesian coordinates can be decomposed into cylindrical harmonics according to (\ref{eq:planetocyl}) as
\begin{align}\label{eq:planetocylAdd}
    \tte_{k} = \sum_{\ell_k\in\RZ} \, \ii^{\,\ell_k} \ \e^{-\ii\,\ell_k \, \vartheta_k} \ \ttc_{k} \ ,
\end{align}
where the corresponding basis states of the  vector space of fields in the $L_\infty$-algebra are
\begin{align*}
    \tte_k(x) &= \tte_{(E_k\,,\,k_x\,,\,k_y\,,\,k_z)}(t,x,y,z) = \e^{-\ii \,E_k\,t +\ii\,k_x\,x+\ii\,k_y\,y+\ii\,k_z\,z} \ ,\\[4pt]
    \ttc_k(x) &= \ttc_{(E_k\,,\,\alpha_{k}\,,\,\ell_k\,,\,k_z)}(t,r,\varphi,z) = J_{\ell_k}(\alpha_k\, r) \ \e^{\,\im\, \ell_k\, \varphi} \ \e^{-\im\, E_k\, t+\im\, k_z\, z} \ ,
\end{align*}
with the coordinate transformation \eqref{eq:radialpolartransf} understood.
The relation (\ref{eq:planetocylAdd}) is the transformation between basis states and it suggests that there is a general transformation between the corresponding objects in both bases. When multiple momenta are involved, we abbreviate subscripts ${}_{k_i}$ by ${}_i$ in what follows.

Let us start with the free propagator of the theory in the plane wave basis (\ref{G2TreeLevel_2p}), which is an elementary contraction 
\begin{align*}
\widetilde G_2(p_1, p_2)^{\swzero} &= -\ii\,\hbar\,\BVL\, \sH \, (\tte^{p_1} \odot \tte^{p_2}) =  \ii\,\hbar \,(2\pi)^4 \ \delta(p_1 + p_2) \ {\widetilde{\sgreen}}(p_1)\\[4pt]
&=  \ii\,\hbar \ {\widetilde{\sgreen}}(p_1) \ \int_{\FR^4}\, \d^4x  \ \tte_{p_1} \cdot \tte_{p_2} \ .
\end{align*}
Decomposing the plane waves under the integral according to (\ref{eq:planetocylAdd}) allows us to perform the integral over the spacetime coordinate $x$ in its cylindrical form as
\begin{align*}
\widetilde G_2(p_1, p_2)^{\swzero} &=  \ii\,\hbar \ {\widetilde{\sgreen}}(p_1) \ \int_{\FR\times\FR} \, \d t\, \d z  \ \int_0^{\infty}\, r\,\d r \ \int_0^{2\pi}\, \d \varphi  \ \sum_{\ell_1, \ell_2\in\RZ} \ \ii^{\,\ell_1+\ell_2} \ \e^{-\ii\,\ell_1\,\vartheta_{1}-\ii\,\ell_2\,\vartheta_{2}} \\
&\hspace{4cm}\times \e^{-\ii\,(\ell_1+\ell_2)\,\varphi} \ \e^{\,\ii\,(E_{1}+E_{2})\,t} \ \e^{-\ii\,(p_{1z}+p_{2z})\,z} \ J_{\ell_1} (\alpha_1\, r) \ J_{\ell_2} (\alpha_2\, r) \ .
\end{align*}

The integrals over the coordinates $(t, \varphi, z)$ trivially yield delta-functions, while the integral over $r$ is the integral of two Bessel functions leading to conservation of radial momentum. Comparing the result with the free propagator of the theory in the cylindrical basis (\ref{G2TreeLevel_2})
proves the transformation relation among free propagators given in different bases:
\begin{align}
\widetilde G_2(p_1, p_2)^{\swzero} = \sum_{\ell_1, \ell_2\in\RZ} \  \ii^{\,\ell_1 + \ell_2} \ 
\e^{-\ii\,\ell_1\, \vartheta_{1} -\ii\,\ell_2\, \vartheta_{2}} \ \widetilde C_2(p_1, p_2)^{\swzero} \ .\label{G0Car_G0Cyl}
\end{align}
Note the implicit dependence on momenta on either side of this transformation: $\widetilde G_2(p_1, p_2)^{\swzero}$ refers to the free propagator in the plane wave basis with momenta $p_i=(E_{i}, p_{ix}, p_{iy}, p_{iz})$ in Cartesian coordinates, or equivalently $p_i=(E_i,\alpha_i,\vartheta_i,p_{iz})$ in cylindrical coordinates after applying \eqref{eq:radialpolartransf}, while $\widetilde C_2(p_1, p_2)^{\swzero}$ refers to the free propagator in the basis of cylindrical harmonics with momenta of the form $p_i=(E_{i}, \alpha_{i}, \ell_{i}, p_{iz})$.

Next we compare the interaction vertices in these two bases. The vertex in the basis of plane waves (\ref{VertexFeynmanPlane_2}) contains the deformed momentum conservation law, expressed as the Dirac distribution in (\ref{StarDelta3_2}):
\begin{align}
V_{\tt plwa}(k_1, k_2, k_3) & = -\frac{g}{3!} \,   (2\pi)^4 \ \delta(k_1+_\star k_2+_\star k_3) = -\frac{g}{3!} \ \int_{\FR^4}\, \d^4x \ \tte_{k_1} \star \tte_{k_2} \star \tte_{k_3} \ .\nn
\end{align}
After decomposing plane waves into  cylindrical harmonics by (\ref{eq:planetocylAdd}), the action of the twist is straightforward since it is diagonal with respect to the basis of cylindrical harmonics. It is easy to extract phase factors and convert the star-product to a pointwise product. The integral over the spacetime coordinate $x$ can again be performed over $(t, \varphi, z)$ resulting in delta-functions, while the integral over $r$ is the integral of three Bessel functions. It is easy to identify the result with the vertex from (\ref{vertex2_2}) given in the basis of cylindrical harmonics, and we arrive at the transformation
\begin{align}
V_{\tt plwa}(k_1, k_2, k_3) &= \sum_{\ell_1, \ell_2,\ell_3\in\RZ} \ \ii^{\,\ell_1 + \ell_2 + \ell_3} \
\e^{-\ii\,\ell_1\, \vartheta_{1}-\ii\,\ell_2\, \vartheta_{2}-\ii\,\ell_3\, \vartheta_{3}} \ V_{\tt cyl}(k_1, k_2, k_3) \ .\label{VCar_VCyl}
\end{align}

We now decompose the one-loop two-point planar diagram (\ref{G2FeynmanPlanarPlane_2}) into cylindrical harmonics. We start from a typical contribution to $\widetilde G_2(p_1, p_2)^{\swone}_{\rm planar}$ of the form
\begin{align*}
&\int_{(\FR^{4})^{\times 3}}\, \frac{\dd^4k_1}{(2\pi)^4} \ \frac{\dd^4k_2}{(2\pi)^4} \ \frac{\dd^4k_3}{(2\pi)^{4}} \ \int_{(\FR^{4})^{\times 3}}\, \frac{\dd^4q_1}{(2\pi)^4} \ \frac{\dd^4q_2}{(2\pi)^4} \ \frac{\dd^4q_3}{(2\pi)^{4}} \ (-\ii\,\hbar)^2\, V_{\tt plwa}(k_1, k_2, k_3) \ V_{\tt plwa}(q_1, q_2, q_3)\\
& \hspace{3cm} \times \BVL\, \sH \, (\tte^{p_1} \odot \tte^{k_1}) \ \BVL\, \sH \, (\tte^{p_2} \odot \tte^{q_1}) \ \BVL\, \sH \, (\tte^{k_2} \odot \tte^{q_3}) \ \BVL\, \sH \, (\tte^{k_3} \odot \tte^{q_2}) \ .
\end{align*}
Implementing the transformations of contractions (\ref{G0Car_G0Cyl}) and vertices (\ref{VCar_VCyl}), after some extensive calculations this results in the diagram (\ref{G2FeynmanPlanarPlane_2}) decomposed in terms of the diagram (\ref{G2FeynmanAdBPlanar_2}) in the basis of cylindrical harmonics:
\begin{align*}
\widetilde G_2(p_1, p_2)^{\swone}_{\rm planar} &= \sum_{\ell_1, \ell_2\in \RZ} \ \ii^{\,\ell_1 + \ell_2} \ 
\e^{-\ii\,\ell_1\, \vartheta_{1}-\ii\,\ell_2\, \vartheta_{2}} \ \widetilde C_2(p_1, p_2)^{\swone}_{\rm planar} \ .
\end{align*}

We have checked explicitly that similar relations hold for the one-loop two-point non-planar diagram as well as for the tree-level three-point function:
\begin{align}
\begin{split}\label{PlaneToCylCorrelationFun}
\widetilde G_2(p_1, p_2)^{\swone}_{\text{non-planar}} &= \sum_{\ell_1, \ell_2\in\RZ} \ \ii^{\,\ell_1 + \ell_2} \
\e^{-\ii\,\ell_1\, \vartheta_{1}-\ii\,\ell_2\, \vartheta_{2}} \ \widetilde C_2(p_1, p_2)^{\swone}_{\text{non-planar}} \ , \\[4pt]
\widetilde G_3(p_1, p_2, p_3)^{\swzero} &= \sum_{\ell_1, \ell_2, \ell_3\in\RZ} \ \ii^{\,\ell_1 + \ell_2 + \ell_3} \ 
\e^{-\ii\,\ell_1\, \vartheta_{1}-\ii\,\ell_2\, \vartheta_{2}-\ii\,\ell_3\, \vartheta_{3}} \ \widetilde C_3(p_1, p_2, p_3)^{\swzero} \ . 
\end{split}
\end{align}
From these results we conjecture that these basis transformations hold in general, that is, for any $n$-point function at any $l$-loop level. This in turn relates the all orders $n$-point correlation functions \eqref{eq:tildeGn} and \eqref{eq:intcorrelationfnMomentum_2} as
\begin{align*}
\widetilde G_n(p_1, \dots, p_n) &= \sum_{\ell_1, \dots, \ell_n\in\RZ} \ \ii^{\,\ell_1 +\dots +\ell_n} \ 
\e^{-\ii\,\ell_1\, \vartheta_{1}-\cdots-\ii\,\ell_n\, \vartheta_{n}} \ \widetilde C_n(p_1, \dots, p_n) \ . 
\end{align*}

As an application of these Fourier series transformations, we can gain some physical insight into how the momentum conservation laws are related between these bases. Let us go back to the tree-level three-point function, which in the bases of plane waves and cylindrical harmonics respectively reads as
\begin{align}
\begin{split}
\widetilde G_3(p_1, p_2, p_3)^{\swzero} &= (\ii\,\hbar)^2\,\frac{g}{3!} \ \widetilde{\sgreen}(p_1) \ \widetilde{\sgreen}(p_2) \ \widetilde{\sgreen}(p_3) \, (2\pi)^4 \ \delta(p_1+_\star p_2+_\star p_3) \ , \\[4pt]
\widetilde C_3(p_1, p_2, p_3)^{\swzero} &= (\ii\,\hbar)^2 \, \frac{g}{3!} \ \widetilde{\sgreen}(p_1) \ \widetilde{\sgreen}(p_2) \ \widetilde{\sgreen}(p_3) \ \e^{\,\frac{\ii\, \lambda}{2} \, (p_{1z}\, \ell_{2} - p_{2z}\, \ell_{1})} \ \delta_{\ell_{1}+\ell_{2}+\ell_{3},0}  \\
&\hspace{1cm} \times (2\pi)^3 \ \delta(E_{1}+E_{2}+E_{3}) \ \delta(p_{1z}+p_{2z}+p_{3z}) \ 
\ttF_{\ell_{1},\ell_{2},\ell_{3}} (\alpha_{1}, \alpha_{2}, \alpha_{3}) \ . \label{C3Example}
\end{split}
\end{align}
These three-point functions are related according to (\ref{PlaneToCylCorrelationFun}).

Consider the special case where $p_1 = (E_{1}, 0,0, p_{1z})$, i.e. with purely axial momentum $\alpha_1=0$.
Inserting the explicit form of $\widetilde C_3(p_1, p_2, p_3)^{\swzero}$ from (\ref{C3Example}) into \eqref{PlaneToCylCorrelationFun} and using \eqref{eq:Fellalphaexplicit} results in
\begin{align*}
\widetilde G_3(p_1, p_2, p_3)^{\swzero}\big|_{\alpha_1=0} & = (\ii\,\hbar)^2\,\frac{g}{3!} \ \widetilde{\sgreen}(p_1) \ \widetilde{\sgreen}(p_2) \ \widetilde{\sgreen}(p_3) \, (2\pi)^4 \ \delta(E_{1}+E_{2}+E_{3}) \ \delta(p_{1z}+p_{2z}+p_{3z}) \nn\\
&\hspace{5cm} \times \frac{\delta(\alpha_{2}-\alpha_{3})}{\alpha_{2}} \ \delta\big(\vartheta_{2} - \vartheta_{3} - \pi + \tfrac{\lambda}{2}\,p_{1z} \big) \ .
\end{align*}
From \eqref{C3Example} we conclude that the deformed conservation of momentum encoded by the Dirac distribution $\delta(p_1+_\star p_2+_\star p_3)\big|_{\alpha_1=0}$  can be understood as a modification of the momentum conservation law in the $(p_x,p_y)$-plane by a shift of the relative polar angle between the momenta by  \smash{$\frac\lambda2\,p_{1z}$}.

\section{Outlook and discussion} \label{Outlook}%

In this paper we applied the algebraic BV quantization method to scalar $\varPhi^3$-theory on $\lambda$-Minkowski space, introducing noncommutativity in two distinct ways. Although both approaches lead to identical Maurer-Cartan functionals, equivalent to the classical action of noncommutative $\varPhi^3$-theory, their quantization schemes differ thus leading to two distinct quantum field theories.

Both theories were expressed in a novel fashion through the spectral decomposition of scalar fields in a basis of cylindrical harmonics, but for different reasons. Braided BV quantization requires the retract maps to be equivariant under the symmetries of the underlying noncommutative spacetime, naturally selecting as basis vectors the $U_\CF\,\fra$-invariant eigenmodes of the Klein-Gordan operator in cylindrical coordinates. In other words, braided BV quantization diagonalises the angular twist by construction. On the other hand, standard BV quantization can be performed in any basis, without restriction, but we argued that the cylindrical harmonic basis is more natural and informative to use. For comparison, standard BV quantization was performed in both the cylindrical harmonic and plane wave bases, with precisely the same results, and the respective correlation functions were explicitly related to one another.

It would be interesting to extend our analysis based on cylindrical harmonics to more intricate quantum field theories on $\lambda$-Minkowski space, and in particular to theories with gauge symmetries. An obvious extension would be to noncommutative quantum electrodynamics. The explicit analysis of quantum correlators for theories with interactions of quartic or higher order, such as scalar $\varPhi^4$-theory and Yang-Mills theory, faces an immediate potential obstacle: to our knowledge, the radial momentum integrals of products of four or more cylindrical Bessel functions of the first kind are not explicitly known in generality. As a toy model to start with, one could investigate  Chern-Simons theory on a spatial section $\FR^3\subset\FR^4$ of $\lambda$-Minkowski space in order to decipher the role of cylindrical harmonics in a theory with non-abelian cubic gauge interactions.

The results of this paper, though model specific, have highlighted at least two general features of noncommutative field theories:
\begin{myitemize}
\item There are two very different quantization schemes. The braided theory is based on a braided $L_\infty$-algebra with explicit noncommutativity. The Braided Wick Theorem compensates for the noncommutative contribution to the interaction vertex, leading to a clear absence of noncommutativity inside loop integrals, and hence of non-planar diagrams. This is in accordance with results from \cite{Bogdanovic:2024jnf} for the case of the Moyal twist deformation. Standard BV quantization, on the other hand, is based on a classical $L_\infty$-algebra with implicit noncommutativity and reproduces the same results obtained by the standard Feynman quantization of noncommutative scalar field theory.
\item Harmonic analysis based on the abelian Lie algebra $\fra=\FR\oplus\frso(2)$ of symmetries of $\lambda$-Minkowski space greatly simplifies calculations and presents results in a form more amenable to physical interpretation. For the Moyal twist, the harmonics are precisely the plane waves associated to the Lie algebra of translations $\FR^d$. Using the same basis for the angular twist leads to deformed momentum conservation laws which partially obscure some features of the quantum field theory, such as the planar equivalence theorem of~\cite{Meier:2023lku} which holds for a broad class of twists based on subalgebras of the Poincar\'e algebra $\mathfrak{iso}(1,3)$. The basis of cylindrical harmonics, on the other hand, encodes noncommutativity through Moyal-like phase factors in interaction vertices and captures the consequences of the violation of standard momentum conservation in a precise fashion. By Frobenius' Theorem, any abelian Drinfel'd twist can be cast into a Moyal-like form through a local change of coordinates, hence investigations of other twist-deformed field theories in this manner are likewise possible, at least in principle. 
\end{myitemize}

\subsubsection*{Acknowledgements}

We thank Stefan Djordjevi{\'c} and Giovanni Landi for helpful discussions. We are grateful to Fedele~Lizzi and Patrizia Vitale for collaboration at the early stages of this project. 
This article is based upon work from COST Actions CaLISTA CA21109 and THEORY-CHALLENGES CA22113 supported by COST (European Cooperation in Science and Technology). Part of this work was carried out during the $5^{\rm th}$ Mini-Symposium on Geometry and Physics at the Rudjer Bo\v{s}kovi{\'c} Institute in Zagreb, Croatia; {\sc MDC} and {\sc RJS} are grateful to the organizers for hospitality and for providing a stimulating environment.
The work of DjB and MDC is supported by Project 451-03-34/2026-03/200162 of the Serbian Ministry of Science,
Technological Development and Innovation. The work of MDC and RJS is partially funded by the Croatian Science Foundation project ``Higher Structures and Symmetries in Gauge and Gravity Theories'' (IP-2024-05-7921).

\appendix

\renewcommand{\theequation}{\Alph{section}.\arabic{equation}}
\setcounter{equation}{0}

\section{Cylindrical Bessel functions of the first kind}
\label{app:Bessel}

In this appendix we summarize some properties of the cylindrical Bessel functions $J_\ell(w)$ for $\ell\in\RZ$ and $w\in\FC$, along with some results on integrals of products of Bessel functions relevant for our calculations in the main text. Further details can be found in \cite{Bailey:1936, Jackson:1972,Gervois:1984,Gradshteyn:2007} and \cite[Chapter~10]{NIST:DLMF}.

The Bessel function $J_\ell(w)$ admits an integral representation along the real line as
\begin{align}
J_\ell (w) = \frac{1}{\pi}\,\int_0^\pi\, {\rm d}\sigma \ \cos (\ell\,\sigma -w\sin\sigma) \ . \label{IntegralRepresentation}
\end{align}
It has the reflection property 
\begin{align*}
J_{-\ell} (w) = (-1)^{\ell}\,J_\ell(w) \ . 
\end{align*}

For $\alpha_i\in(0,\infty)$, the Bessel functions obey the orthogonality relation
\begin{align}
\int_0^\infty\,r\,\dd r \ J_{\ell} (\alpha_1\, r) \, J_{-\ell} (\alpha_2\, r)  = \frac{(-1)^{\ell}}{\alpha_1} \ \delta (\alpha_1 - \alpha_2) \ . \label{2Bessel} 
\end{align}

The integration of the product of three Bessel functions is given by~\cite[eq.~(40)]{Jackson:1972}
\begin{align}
\begin{split}
& \int_0^\infty\,r\,\dd r  \  J_{\ell_1} (\alpha_1\, r) \ J_{\ell_2} (\alpha_2\, r) \ J_{-\ell_1-\ell_2} (\alpha_3\, r) \\[4pt]
& \hspace{4cm} =  \int_{[0,2\pi)^{\times3}}\, \frac{{\rm d}\vartheta_1\, {\rm d}\vartheta_2\, {\rm d}\vartheta_3}{(2\pi)^3} \ \delta(\vec{\alpha}_1 + \vec{\alpha}_2 + \vec{\alpha}_3) \
\e^{\,\im\,\ell_1\,(\vartheta_1-\vartheta_3)+\im\,\ell_2\,(\vartheta_2-\vartheta_3)} \ , \label{3Bessel1}
\end{split}
\end{align}
which can be explicitly evaluated to
\begin{align}
\begin{split}
& \int_0^\infty\,r\,\dd r \ J_{\ell_1} (\alpha_1\, r) \ J_{\ell_2} (\alpha_2\, r) \ J_{-\ell_1-\ell_2} (\alpha_3\, r) \\[4pt]
& \qquad = \begin{cases}
 \ 0 \ , &\quad \alpha_3> \alpha_1+\alpha_2 \quad \text{or} \quad 0<\alpha_3<|\alpha_1-\alpha_2| \ , \\[4pt]
 \ \displaystyle (-1)^{\ell_1+\ell_2} \ \frac{\cos(\ell_1\,\upsilon_2 - \ell_2\,\upsilon_1)}{\pi\, \alpha_1\,\alpha_2\sin(\upsilon_3)} \ ,& \quad |\alpha_1-\alpha_2|<\alpha_3 < \alpha_1 + \alpha_2 \ ,\> \upsilon_1+\upsilon_2+\upsilon_3 = \pi \ . 
\end{cases} \label{3Bessel}
\end{split}
\end{align}

The notation used in the formulas \eqref{3Bessel1} and \eqref{3Bessel}, and in particular the definition of the angles $\vartheta_i$ and $\upsilon_i$, may be represented pictorially as
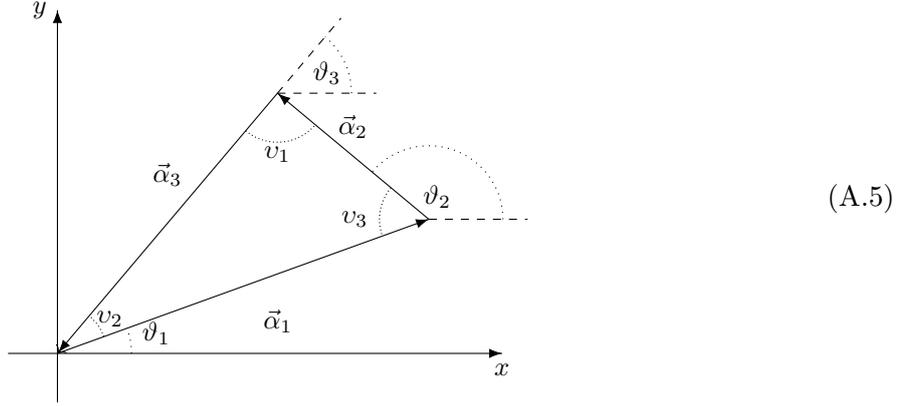
\begin{align}
\begin{split}
\small
    \begin{tikzpicture}[scale=0.65, baseline]
        \coordinate (O) at (0,0);
        \coordinate (A) at ($(O)+(20:8)$);
        \coordinate (B) at ($(A)+(140:4)$);
        \coordinate (S1) at ($0.5*(A)$);
        \coordinate (S2) at ($0.5*(A)+0.5*(B)$);
        \coordinate (S3) at ($0.5*(B)$);
        \draw[decoration={markings, mark=at position 1 with {\arrow[scale=1]{Latex}} }, postaction={decorate}] ($(O) +(180:1)$) -- +(0:10) node[below=0.5]{$x$};
        \draw[decoration={markings, mark=at position 1 with {\arrow[scale=1]{Latex}} }, postaction={decorate}] ($(O)+(-90:1)$) -- +(90:8) node[left=0.5]{$y$};
\draw[decoration={markings, mark=at position 1 with {\arrow[scale=1]{Latex}} }, postaction={decorate}] (O) -- (A);
        \draw[decoration={markings, mark=at position 1 with {\arrow[scale=1]{Latex}} }, postaction={decorate}] (A) -- (B);
        \draw[decoration={markings, mark=at position 1 with {\arrow[scale=1]{Latex}} }, postaction={decorate},->] (B) -- (O);
\draw ($(S1)+(-45:1)$) node{$\Vec{\alpha}_{1}$};
        \draw ($(S2)+(90:0.6)$) node{$\Vec{\alpha}_{2}$};
        \draw ($(S3)+(90:1)$) node{$\Vec{\alpha}_{3}$};
        \draw ($(O)+(0:2)$) node[above]{$\vartheta_1$};
        \draw ($(O)+(45:1.5)$) node[below]{$\upsilon_2$};
        \draw ($(A)+(70:0.5)$) node{$\vartheta_2$};
        \draw ($(A)+(180:1.5)$) node{$\upsilon_3$};
        \draw ($(B)+(0:1)$) node[above]{$\vartheta_3$};
        \draw ($(B)+(-90:1.25)$) node{$\upsilon_1$};
\draw[dashed] (A) -- +(0:2);
        \draw[dotted] (A)+(0:1.5) arc [start angle=0, end angle=140, radius=1.5];
        \draw[densely dotted] (A)+(140:1) arc [start angle=140, end angle=200, radius=1];
        \draw[dashed] (B) -- +(0:2);
        \draw[dashed] (B) -- +(50:2);
        \draw[dotted] (B)+(0:1.5) arc [start angle=0, end angle=50, radius=1.5];
        \draw[densely dotted] (B)+(230:1) arc [start angle=230, end angle=320, radius=1];
        \draw[dotted] (O)+(0:1.5) arc [start angle=0, end angle=20, radius=1.5];
        \draw[densely dotted] (O)+(20:1) arc [start angle=20, end angle=50, radius=1];
    \end{tikzpicture}
    \label{fig:Bessel1}
\normalsize
\end{split}
\end{align}
Here we have chosen auxiliary $x$ and $y$ axes for illustration purposes only. None of our results depends on that choice since the angular twist respects cylindrical symmetry.

In \eqref{fig:Bessel1} we have adopted the following conversion. By promoting $\alpha > 0$ to the vector $\vec{\alpha} = \alpha\, \vec{e}_r$, where $\vec e_r$ is the unit radial vector in the $(x,y)$-plane,  we can satisfy the condition under which the integral in \eqref{3Bessel} is non-trivial by the choice 
\begin{equation}
    \vec{\alpha}_1 + \vec{\alpha}_2 + \vec{\alpha}_3 = \vec 0 \ . \label{3momenta}
\end{equation}
The constraint (\ref{3momenta}) represents the triangle in \eqref{fig:Bessel1}, in which the trigonometric laws of sines and cosines apply. In particular, from the area of the triangle we have
\begin{align}\nn
    \alpha_1\,\alpha_2 \sin(\upsilon_3) = \alpha_1\,\alpha_3 \sin(\upsilon_2) = \alpha_2\,\alpha_3 \sin(\upsilon_1)  \ ,
\end{align}
along with
\begin{align*}
    2\,\alpha_1\,\alpha_2\cos(\upsilon_3) = \alpha_1^2+\alpha_2^2-\alpha_3^2 \ ,
\end{align*}
as well as two analogous cosine laws corresponding to cyclic permutations of the indices $(123)$.

The integral \eqref{3Bessel} is non-zero only when all $\alpha_i\neq0$. When one sets $\alpha_3 = 0$, one can use \eqref{IntegralRepresentation} to deduce
\begin{equation}
J_{\ell} (0) = \delta_{\ell,0} \ ,\label{Jzerozero}
\end{equation}
which implies $\ell_1+\ell_2 = 0$. Then the integral \eqref{2Bessel} yields
\begin{align}\label{2BesselAdd}
    \int_0^\infty\,r\,\dd r \ J_{\ell_1} (\alpha_1\, r) \ J_{\ell_{2}} (\alpha_{2}\, r) \ J_{-\ell_1-\ell_2} (0) = \delta_{\ell_1+\ell_2,0} \ \frac{(-1)^{\ell_1}}{\alpha_1} \ \delta(\alpha_1-\alpha_2) \ .
\end{align}

\bigskip

\bibliographystyle{ourstyle}  
\bibliography{LMRef.bib}

\end{document}